\documentclass{emulateapj}                                                                                   
\usepackage{color}
\usepackage{amsmath}
\usepackage{natbib}
\usepackage{graphicx}
\usepackage{epstopdf}
\usepackage{units}
\usepackage{relsize}
\usepackage{enumitem}  
\usepackage{aas_macros}                                                                             
              
\def\hmpc{h^{-1}{\rm Mpc}}
\def\hkpc{h^{-1}{\rm kpc}}
\def\kpch{{\rm kpc}/h}
\def\msun{{\rm ~M}_{\odot}}

\def\wtheta{{\omega(\theta)}}
\def\Mmin{M_\mathrm{min}}
\def\fgal{f_\mathrm{gal}}

\begin{document}

\title{ The Spatial Distribution of Satellite Galaxies Within Halos: \\ Measuring the Very Small Scale Angular Clustering of SDSS Galaxies}

\author{Jennifer A. Piscionere\altaffilmark{1}, Andreas A. Berlind \altaffilmark{1}, Cameron K. McBride, \altaffilmark{2}, Rom\'{a}n Scoccimarro \altaffilmark{3}}
 \affil{
     $^{1}$ Department of Physics and Astronomy, Vanderbilt University, Nashville, TN 37235\\
     $^{2}$ Center for Astrophysics, Harvard University, Cambridge, MA 00000\\
     $^{3}$ Center for Cosmology and Particle Physics, Department of Physics, New York University, New York, NY 10003\\
     j.piscionere@vanderbilt.edu}

\begin{abstract}
We measure the angular clustering of galaxies from the Sloan Digital Sky Survey Data Release 7 in order to probe the spatial distribution of satellite galaxies within their dark matter halos. Specifically, we measure the angular correlation function on very small scales ($7-320\arcsec$) in a range of luminosity threshold samples (absolute $r$-band magnitudes of -18 up to -21) that are constructed from the subset of SDSS that has been spectroscopically observed more than once (the so-called plate overlap region). We choose to measure angular clustering in this reduced survey footprint in order to minimize the effects of fiber collision incompleteness, which are otherwise substantial on these small scales, and we discuss the possible impact that fiber collisions have on our measurements. We model our clustering measurements using a fully numerical halo model that populates dark matter halos in N-body simulations to create realistic mock galaxy catalogs. The model has free parameters that specify both the number and spatial distribution of galaxies within their host halos. We adopt a flexible density profile for the spatial distribution of satellite galaxies that is similar to the dark matter Navarro-Frenk-White (NFW) profile, except that the inner slope is allowed to vary. We find that the angular clustering of our most luminous samples ($M_r<$ -20 and -21) suggests that luminous satellite galaxies have substantially steeper inner density profiles than NFW. Lower luminosity samples are less constraining, however, and are consistent with satellite galaxies having shallow density profiles. Our results confirm the findings of \cite{Watson2012} while using different clustering measurements and modeling methodology.
\end{abstract}


%================================================
\section{Introduction}

%Large Scales
One of the best statistical tools we have for an understanding of the galaxy distribution in the universe is the two-point correlation function of galaxies. On large scales (approximately greater than $10\hmpc$), galaxies are simple tracers of the underlying matter density field and so the correlation function (or its Fourier equivalent, the power spectrum) can be used to probe the nature of matter fluctuations, and thus yield constraints on cosmological parameters \citep[e.g.,][]{Tegmark2004,Tegmark2004WMAP}. At a scale of $\sim 100\hmpc$, the Baryon Acoustic Oscillation (BAO) feature in the correlation function also provides strong cosmological constraints \citep[e.g.,][]{Eisenstein2005}. 

%Small/Intermediate Scales <10Mpc?
On scales smaller than $\sim 10\hmpc$, the galaxy correlation function encodes information about the detailed relationship between the spatial distribution of galaxies and the underlying dark matter, which is substantially more complex than on large scales. Adopting the assumption that all galaxies live within dark matter halos, the halo model provides a useful roadmap for interpreting galaxy clustering on these scales. In the halo model framework, the clustering of galaxies can be calculated from statistical properties of halos, such as their abundance, clustering and internal structure, combined with parameterized relations that describe how galaxies occupy halos. This latter part is referred to as the Halo Occupation Distribution (HOD) and it typically specifies the number of galaxies as a function of halo mass, together with an assumption for the their spatial distribution within halos \citep[e.g.,][]{Peacock2000,Scoccimarro2001,Berlind2002HOD,CooraySheth2002,Berlind2003,Kravtsov2004,Zheng2005}. Several studies have used the measured galaxy correlation function on scales of $\sim 0.1-10\hmpc$ to constrain the HOD and thus illuminate the nature of the connection between galaxies and dark matter \citep[e.g.,][and references therein]{Zehavi2005,Zehavi2011,ZhengCoilZehavi,Guo2014}. 

%Very small scales
On very small scales, well within $0.5\hmpc$, the typical size of halos that host bright galaxy pairs, the shape of the correlation function is primarily dictated by the spatial distribution of galaxies in each halo \citep[e.g.,][]{Berlind2002HOD,Zehavi2005}. Most studies adopt a simple model whereby the first ``central" galaxy in each halo lives at the halo center, and subsequent ``satellite" galaxies trace the density distribution of the dark matter. Specifically, satellite galaxies are usually assumed to follow a Navarro-Frenk-White \citep[NFW;][]{NFW} profile, which does a good job of describing the density profiles of halos in pure dark matter N-body simulations. This assumption is theoretically motivated \citep[e.g.,][]{Berlind2003} and it works well in explaining the observed shape of the correlation function on small scales.

%Masjedi & Watson:
The first evidence from galaxy clustering that satellite galaxies might not actually trace mass within halos came from \cite{Masjedi2006} who pushed the measurement of the galaxy correlation function down to scales of $10\hkpc$.  Using a sample of Luminous Red Galaxies \citep[LRGs;][]{Eisenstein2001} selected from the Sloan Digital Sky Suvey \citep[SDSS;][]{York2000}, \cite{Masjedi2006} found that the correlation function of LRGs at the smallest scales ($\lesssim 30\hkpc$) was under-predicted by the \cite{zehavi2005LRG} HOD model that had successfully fit the clustering of the same galaxies at larger scales. Specifically, the HOD model predicted a $r^{-1}$ slope for the correlation function at the smallest scales (which comes from the inner slope of the NFW profile), whereas \cite{Masjedi2006} found a much steeper $r^{-2}$ slope. \cite{Watson2010} explored this discrepancy in detail by fitting the \cite{Masjedi2006} correlation function measurements with a HOD model that relaxed the assumption that satellite galaxies follow a NFW profile.  Instead, they adopted a more flexible profile where the inner slope was allowed to vary.  \cite{Watson2010} were able to obtain a good fit to the LRG clustering for a satellite galaxy profile with an $r^{-2}$ inner slope while ruling out the NFW profile at high significance.

%Watson2, Guo
\cite{Watson2012} extended this work to a wider range of galaxy luminosities. They fit their flexible HOD model to measurements of the projected correlation function, $w_p(r_p)$, in several SDSS luminosity samples, ranging from absolute $r$-band magnitude of -18 on the faint end, to LRGs on the bright end. These measurements were made by \cite{Jiang2012} using the same methods as \cite{Masjedi2006} for pushing to very small scales. \cite{Watson2012} found a clear luminosity trend whereby the clustering of galaxy samples with $M_r<$ -20 and brighter demanded steeper density profiles for satellite galaxies than NFW, whereas lower luminosity samples were consistent with NFW satellite profiles. \cite{Guo2014} adopted the same flexible density profile when modeling the clustering of galaxies in the SDSS III \citep{Eisenstein2011} Baryon
Oscillation Spectroscopic Survey \citep[BOSS;][]{Dawson2013}, and also found a significant departure from NFW, albeit only for the reddest galaxies in that survey. Unfortunately, it is difficult to directly compare these results with those of \cite{Watson2012} because of the substantially different sample selections.
Using a different technique that does not involve correlation functions, \cite{Tal2012b} found that satellite galaxies around LRGs deviate from NFW at very small scales, in agreement with \cite{Watson2010}. On the other hand, \cite{GuoCole2012} used a similar technique to find that satellite galaxies have density profiles that are consistent with NFW. Deep imaging of satellites around luminous Early-type galaxies at intermediate redshifts have shown an isothermal profile \citep{Nierenberg2011} with no dependence on host mass \citep{Nierenberg2012}.

%Systematics Paragraph
Measurements of the galaxy correlation function on such small scales suffer from two potentially severe systematic errors. First, two bright galaxies that are only separated by $\sim10-30\hkpc$ are likely in the process of merging and will have overlapping light profiles. It can be difficult to accurately de-blend the observed light into two separate components and a sufficiently large error in the assigned magnitude of either galaxy can cause the pair to either enter or drop out of a luminosity selected sample. Second, in surveys that use fiber-fed multi-object spectrographs, it is not possible to obtain spectra of both galaxies that are separated by less than the physical diameter of the fibers. In the SDSS, these ``fiber collisions" enter at an angular scale of $55\arcsec$ \citep{BlantonTiling}. At the typical redshifts of SDSS galaxies, this corresponds to a much larger physical scale than $30\hkpc$. About a third of these collided galaxy pairs are recovered in the SDSS because part of the survey footprint is observed (``tiled") more than once. However, the spatial distribution of this overlap region is very complex. Incompleteness due to fiber collisions affects the correlation function the most on the smallest scales, but the $55\arcsec$ angular scale translates into many different length scales in real and projected space, so even large scales are affected. Various methods have been used to correct for fiber collisions. The simplest method is to assign collided galaxies the redshifts of their nearest neighbors. This works well on large scales, but not small scales. \cite{Masjedi2006} and \cite{Jiang2012} used an estimator for $w_p(r_p)$ that corrects for fiber collision incompleteness statistically. \cite{Guo2014} used a different method that essentially only considers galaxy pairs in overlap regions \citep{Guo2012}. It is important to correctly account for these systematic effects before drawing any conclusions about the inner density profile of satellite galaxies.

%why this is interesting
It should not necessarily come as a surprise that satellite galaxies may not be perfect tracers of dark matter. The spatial distribution of satellite galaxies can be affected both by dynamical mechanisms, such as dynamical friction and tidal stripping of stars due to the host halo potential, and by baryonic processes, such as quenching of satellite star formation. A detection of a departure from the dark matter profile in the satellite density profile can thus serve as a probe of these processes. Theoretical predictions of the satellite galaxy density profile at such small scales are difficult to make because it is challenging to resolve massive distinct satellite halos (i.e., subhalos) so close to the center of a larger host halo. Nevertheless, both pure N-body and hydrodynamic simulations are now achieving the resolutions and volumes necessary to compare with SDSS data \citep[e.g.,][]{Pujol2014,Genel2014}. 

%what we will do
In this paper, we test the validity of the \cite{Watson2012} results using the same galaxy selection, but an entirely different methodology. First, we measure the angular correlation function $\wtheta$, instead of the projected function $w_p(r_p)$. In general, $\wtheta$ is a powerful tool for two-dimensional galaxy surveys (see \citealt{Crocce2011} and references therein).  It has been employed to measure the galaxy clustering in the early data release of the SDSS \citep{Connolly2002,Scranton2002,Infante2002,Budavari2003}, as well many other galaxy surveys \citep[e.g.,][]{Groth1977,McCracken2001,Maller2005}. The angular function is less sensitive to fiber collisions because the fiber incompleteness enters at a fixed scale and thus does not contaminate larger scales.  Moreover, we restrict our samples to survey overlap regions, which reduces the effects of fiber collisions even more.  Second, we improve on the HOD modeling by switching to an accurate and fully numerical way of computing clustering predictions, instead of the quick and approximate analytic method that was used in \cite{Watson2012}.

%layout of paper
The description of our data samples appears in \S\ref{Data_Sample}. The $\wtheta$ measurements, along with power-law fits, are described in \S\ref{Wtheta}. The description of our modified density profile HOD model is in \S\ref{model}, with results of the model fits presented in \S\ref{results}. In \S\ref{discussion} we summarize our results and discuss their implications. Finally, we discuss fiber collision incompleteness in the Appendix.Throughout this paper, we assume a standard $\Lambda$CDM cosmology in concordance with the best fit WMAP5 parameters.

%================================================
\section{Data Sample}
\label{Data_Sample}

%SDSS spectroscopic sample
Measuring angular correlations does not usually require galaxies with measured redshifts. However, we wish to constrain the density profile of satellite galaxies within their halos for different luminosity samples so that we can test the \cite{Watson2012} results. We therefore need volume-limited samples built from a spectroscopic sample. We use data from the SDSS Data Release 7 \citep[DR7;][]{SDSSDR7}. Specifically, we use the large-scale structure samples from the NYU Value Added Galaxy Catalog (NYU-VAGC; \citealt{vagc}), that were built from the SDSS main galaxy sample \citep{Strauss2002}. The main spectroscopic galaxy sample is approximately complete down to an apparent $r$-band Petrosian magnitude limit of $<17.77$.  However, we have cut our sample back to $r<17.6$ so that it is complete down to that magnitude limit across the sky. Galaxy absolute magnitudes are k-corrected \citep{blantonKCorrection} to rest-frame magnitudes at redshift $z=0.1$. 

\begin{figure}
\centering
\includegraphics[width=3.5in,height=3.5in]{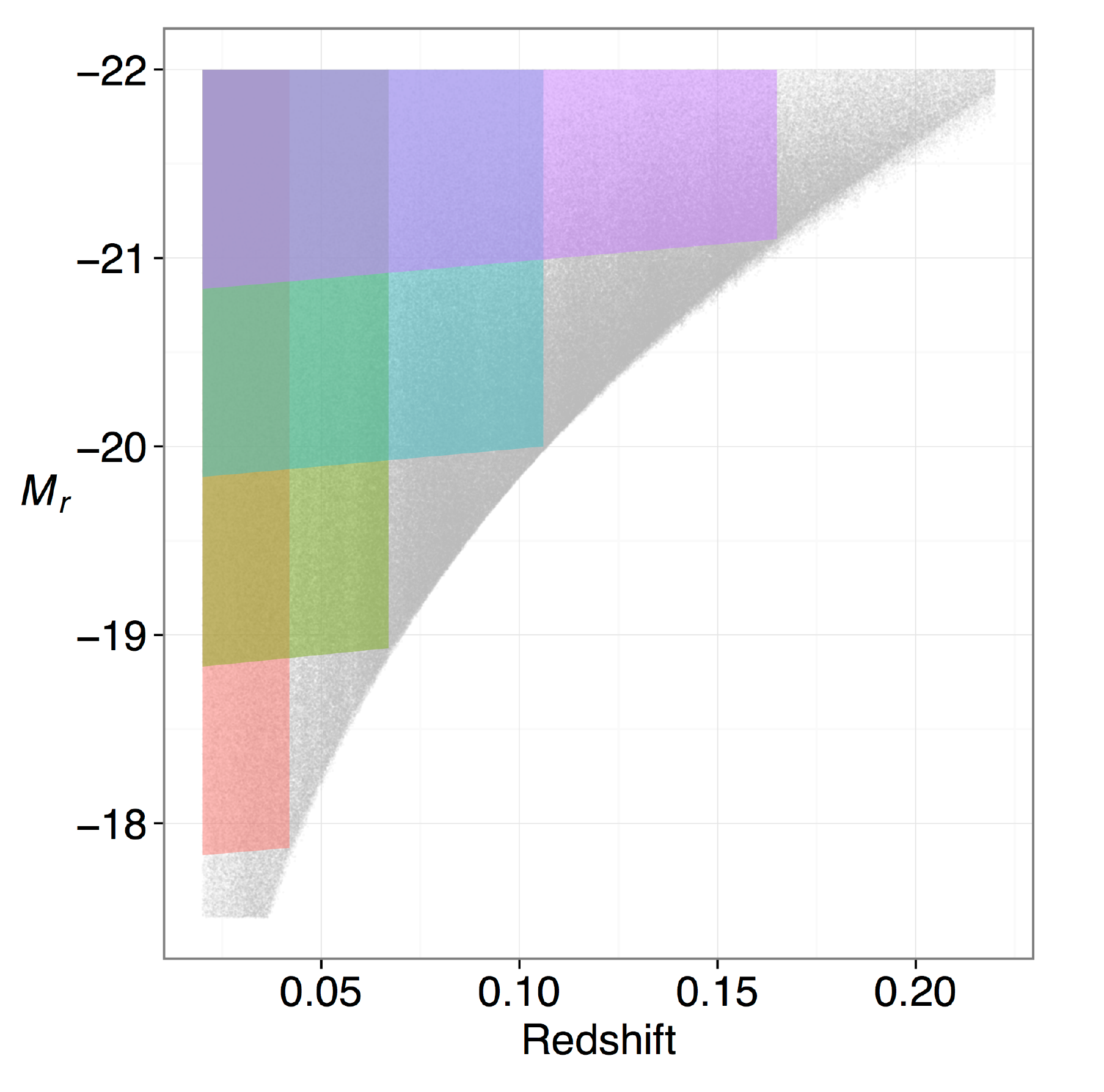}
\caption{The four volume-limited samples that we use in this study, shown in absolute $r$-band magnitude vs. redshift. Light grey points show the full flux-limited sample from which the volume-limited samples were selected. The absolute magnitude limits of the four samples evolve slightly with redshift to account for passive luminosity evolution in the galaxy population.
}
\label{vollim}
\end{figure}

%volume-limited samples
We construct four volume-limited samples that are complete down to absolute $r$-band magnitude limits of -18, -19, -20, and -21. When constructing the volume-limited samples, we adopt corrections for passive luminosity evolution \citep{BlantonEvolCorrection}, which results in slightly evolving absolute magnitude limits as a function of redshift (the magnitude limits listed above apply at $z=0.1$). The four volume-limited galaxy samples are shown in Figure~\ref{vollim} and their redshift limits and sizes are summarized in Table~\ref{gal_summary}. 

%-------------------------------
%-------------------------------
%-------------------------------
\begin{deluxetable}{lcccccc}
\tablecolumns{7}
\tabletypesize{\scriptsize}
\tablewidth{0pt}
\tablecaption{Volume-limited Samples and Power Law Fits}
\tablehead{
& & & & Collision & Median & \\ 
$M^{\mathrm{lim}}_r$ & $z_{\mathrm{min}}$ & $z_{\mathrm{max}}$ & $N_{\mathrm{gal}}$ & Scale & Slope & $\nicefrac{\chi^2}{\mathrm{dof}}$ \vspace{0.05cm} \\
& & & & ($\kpch$) & & \vspace{-0.15cm}\\ 
}
\startdata
\vspace{0.2cm}
-18 & 0.02 & 0.042 & 18690 & 25.7 & $-0.70\pm 0.05$ & 1.45 \\
\vspace{0.2cm}
-19 & 0.02 & 0.067 & 41515 & 39.9 & $-0.77\pm 0.02$ & 0.70 \\
\vspace{0.2cm}
-20 & 0.02 & 0.106 & 67108 & 59.9 & $-0.74\pm 0.02$ &1.06 \\
-21 & 0.02 & 0.165 & 43528 & 89.1 & $-0.92\pm 0.02$ &0.779 \\
\vspace{-0.15cm}
\enddata
\tablecomments{The table shows the absolute magnitude and redshift limits of each sample, the number of galaxies, the physical scale of fiber collisions at the median redshift of the sample, and the median slope and best-fit $\chi^2$ from fitting a power law to the angular correlation function.}
\label{gal_summary}
\end{deluxetable}
%-------------------------------

%fiber collisions are a problem
The galaxy redshift sample has an incompleteness due to the mechanical restriction that spectroscopic fibers cannot be placed closer to each other than their own 
thickness. This fiber collision constraint makes it impossible to obtain redshifts for both galaxies in pairs that are closer than $55\arcsec$ on the sky.  In the case of a 
conflict, the target selection algorithm randomly chooses which galaxy gets a fiber \citep{Strauss2002}. Spectroscopic plate overlaps alleviate this problem to some extent, but fiber collisions still account for a $\sim 6\%$ incompleteness in the main galaxy sample. On the very small scales that we are considering in this paper, fiber collision incompleteness is severe. The $55\arcsec$ angular scale translates to physical scales of $25-90\hkpc$ at the median depths of our four samples, which is right in the interesting region we wish to study.

%-------------------------------
\begin{figure*}
\centering
\includegraphics[width=7in]{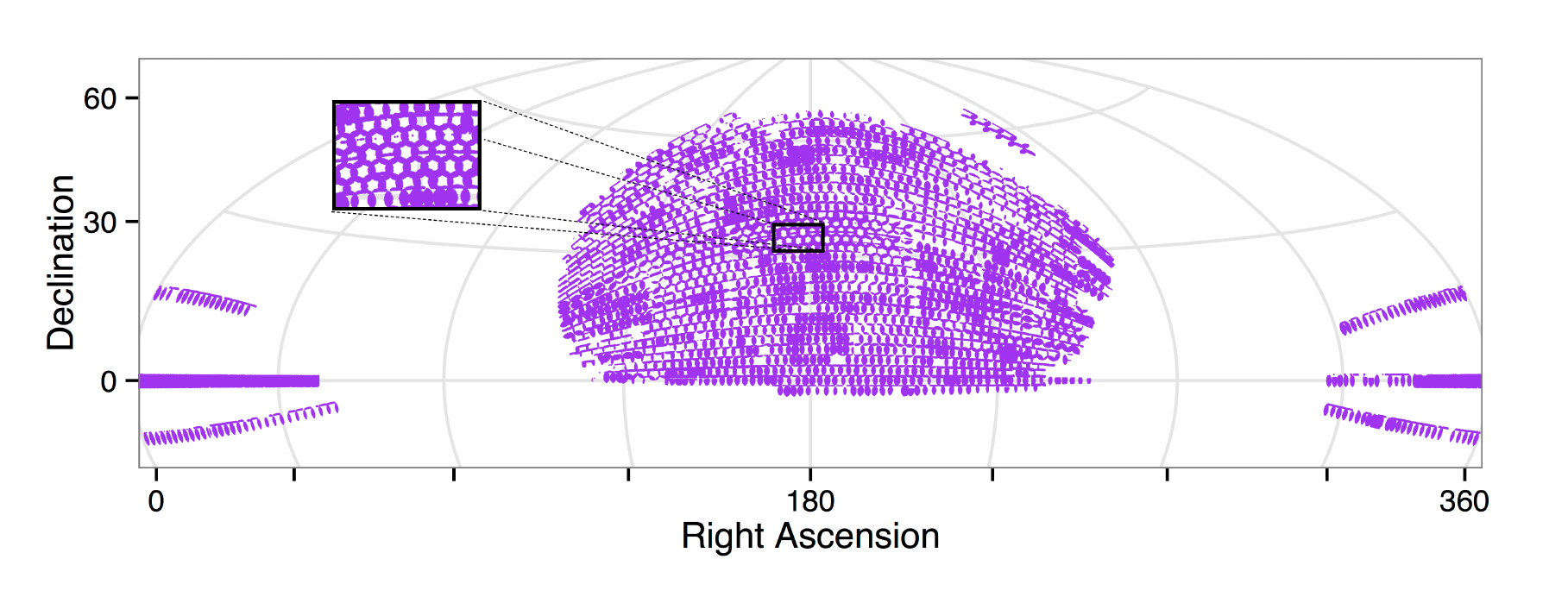}
\caption{The footprint on the sky (Hammer projection) of the SDSS `overlap' sample that we use in this paper. The sample consists only of regions that were spectroscopically observed more than once. The area of this footprint is roughly 40\% of the full SDSS DR7 footprint.}
\label{overlap}
\end{figure*}
%-------------------------------

%how to correct for fiber collisions
A commonly used correction for fiber collisions in galaxy clustering studies is to assign fiber collided galaxies the redshift of the galaxy they collided with \citep[i.e., the ``nearest neighbor correction";][]{Zehavi2002}. This correction recovers the true correlation function on large scales, but it performs poorly on small scales. \cite{Masjedi2006} and \cite{Jiang2012} addressed this problem by proposing a new estimator for the projected correlation function. Instead of computing an autocorrelation function of spectroscopic galaxies, they computed a cross-correlation between spectroscopic galaxies and all spectroscopic targets from the imaging survey. For each pair, the imaging galaxy was placed at the same redshift as the spectroscopic galaxy. This procedure recovers all the fiber collided pairs, but it also includes an artificial signal from pairs that are actually uncorrelated. The uncorrelated pairs are then statistically removed from the correlation function using a random catalog. In this paper, we adopt a different approach. We apply the nearest neighbor correction to recover collided galaxies without redshifts, we construct our samples, and then we measure the angular correlation function of galaxies, $\wtheta$. The angular function is significantly less sensitive to errors in the assigned redshifts than the projected function $w_p(r_p)$ because the angular scale $\theta$ is not affected by these errors, whereas the physical scale $r_p$ is. Errors in the nearest neighbor correction only affect $\wtheta$ if they cause galaxies to enter or drop out of the volume-limited sample. For most collision pairs, the nearest neighbor correction does not result in the gain or loss of the pair in the sample. This only happens in special cases. For example, when a collision pair straddles the outer redshift limit of a particular volume-limited sample, if the more distant galaxy of the pair did not get a redshift due to the collision, the nearest neighbor correction will bring it into the sample and thus result in a new small-scale pair contributing to $\wtheta$. Alternatively, if the higher redshift galaxy of a collision pair is close to the luminosity limit of the sample and did not get a redshift due to the collision, the nearest neighbor correction could make it exit the sample. This would result in a loss of a small scale pair contributing to $\wtheta$.

%overlap footprint
The SDSS DR7 sample covers an area on the sky of approximately 8000 square degrees. However, to minimize the errors due to fiber collisions discussed above, we restrict the sample to regions on the sky that have been spectroscopically observed more than once (the so called ``plate overlap" regions) as part of the tiling process \citep{BlantonTiling}. In these regions, which cover about 40\% of the full SDSS footprint, the vast majority of collided galaxies have been recovered. However, we note that a region that has been tiled twice can only recover close $pairs$ of galaxies. In order to measure the redshifts of close $triplets$, a region would have to be tiled thrice. This continues on to higher groups, which represent a small number of the collision groups, but a non-negligible fraction of pairs. The effects of fiber collisions are thus not completely removed from our analysis and we revisit this issue in \S\ref{Wtheta} and in the Appendix. The total area of our sample is 3300 square degrees and we refer to it as the `overlap' sample throughout this paper. We show the sample footprint in Figure~\ref{overlap}.

%================================================
\section{Angular Correlation Function}
\label{Wtheta}

\subsection{Measuring $\wtheta$}
\label{measurement}

%estimator
We measure $\wtheta$ using the \cite{LandySzalay1993} estimator
\begin{equation}
\label{lsestimator}
w(\theta) = \frac{DD -2DR+ RR}{RR},
\end{equation}
\noindent where $DD$, $DR$ and $RR$ are the correctly normalized number of data-data, data-random and random-random pairs in each bin of angular separation $\theta$. We construct a random sample that has the same overlap geometry as the data sample and a size such that the amount of shot noise in the inner bins is not dominated by $RR$ or $DR$.  

%jackknife
In order to estimate errors and measure the covariance matrix, we separate the footprint into 100 jackknife samples that represent approximately equal area sections on the sky. For each jackknife sample $k$, we measure the angular correlation function $\omega^k(\theta)$. The covariance matrix can then be computed as
\begin{equation}
\label{covar_jack}
C_{ij} = \frac{N-1}{N} \sum_{k=1}^{N} (\omega^k_i - \bar{\omega_i})(\omega^k_j - \bar{\omega_j}),
\end{equation}
where $C_{ij}$ is the covariance between angular bins $i$ and $j$, and $\bar{\omega_i}$ is the mean of correlation function measurements in angular bin $i$ computed from the $N$ jackknife samples. We will use the full covariance matrix to fit models to our measurements since neighboring data points in the angular correlation function are highly correlated \citep{Connolly2002}.  

%Stomp
The measurement of $\wtheta$ is done using \texttt{STOMP}, a C++ library platform for doing fast spatial statistics on arbitrary spherical geometries using 10s of millions of points\footnote{http://code.google.com/p/astro-stomp/}.  The 100 jackknife samples of equal area on the sky are made using the \texttt{STOMP} libraries.  
 
\subsection{Data Results}

%-------------------------------
\begin{figure*}
\centering
\includegraphics[width=5in,height=5in]{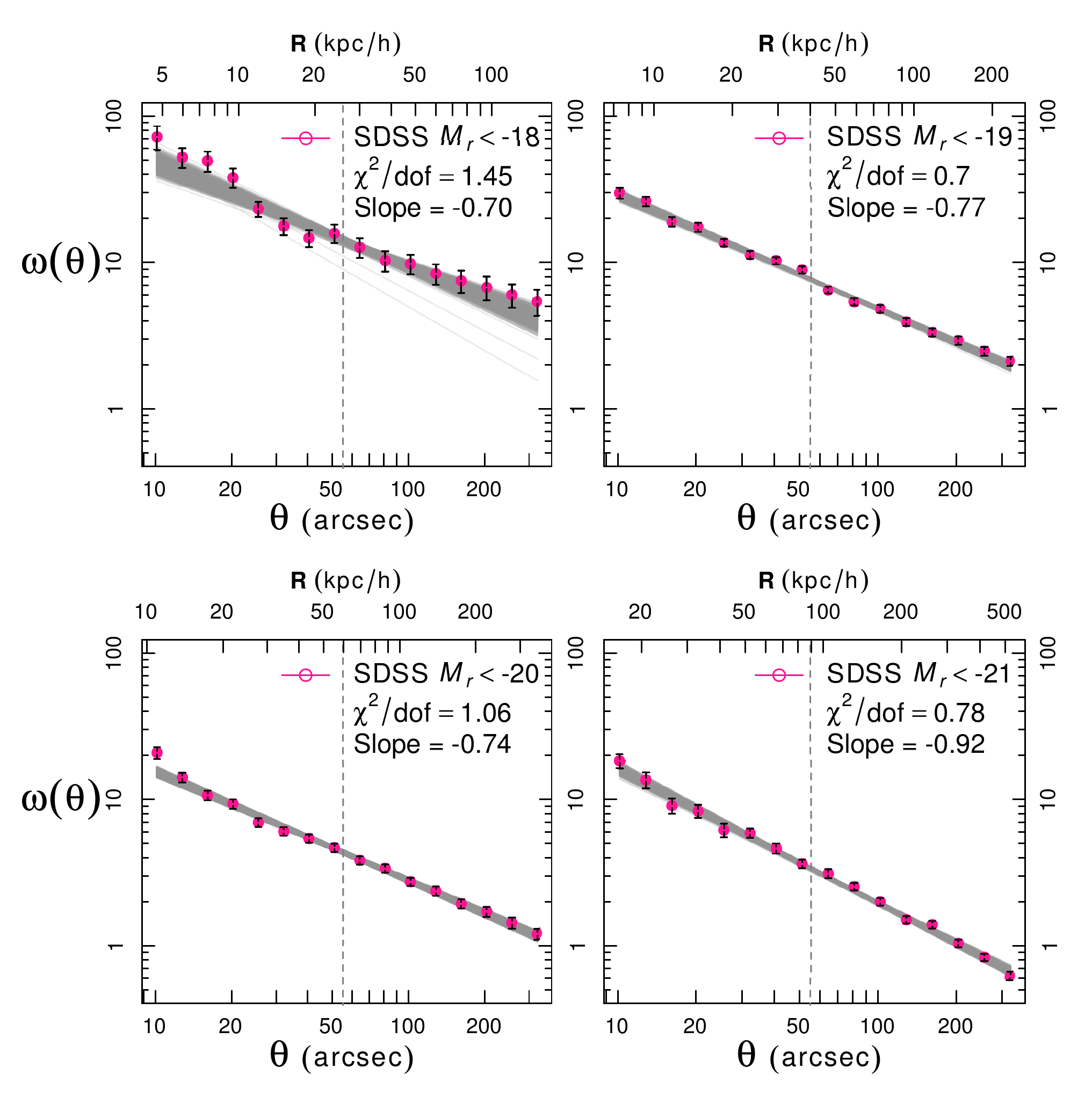}
\caption{The angular correlation function of SDSS galaxies in four volume-limited samples, along with their power-law fits. Each panel shows results for a specific volume-limited sample, as described in \S\ref{Data_Sample}.  Points show the measurements, and error bars are estimated from jackknife resampling of the data on the sky. The bottom axis of each panel shows the angular scale in units of arcsec and the top axis shows the corresponding physical scale at the median redshift of each sample. The vertical dashed line in each panel denotes the fiber collision scale of 55$\arcsec$. The gray band shows a selection of power law models that are randomly drawn from the best-fitting 68\% of models in the MCMC chain. The median value of the slope and the goodness of fit are listed in each panel.}
\label{4-plot}
\end{figure*}
%-------------------------------

%Wtheta scales covered
Figure~\ref{4-plot} presents our measurements of the angular correlation function $\wtheta$, for the four volume-limited samples described in \S\ref{Data_Sample} in the range $7\arcsec< \theta < 320\arcsec$.  We choose this range of scales because on smaller scales photometric deblending effects are expected to be severe, while larger scales no longer probe the clustering of galaxies within a single dark matter halo. \cite{Masjedi2006} quantified the effects of photometric deblending on the correlation function for LRGs by adding artificial galaxy pairs into the raw SDSS images and studying how well the photometric pipeline recovered the light of each galaxy. They found that the clustering of LRGs is significantly overestimated on scales less than $20\hkpc$ due to deblending errors, while larger scales are mostly unaffected. Since the physical sizes of galaxies decrease rapidly with decreasing luminosity, \cite{Jiang2012} calculated that it is safe to ignore photometric deblending effects for the lower luminosity samples and scales we consider here. The physical scales corresponding to these angular scales for the median redshift of each sample are shown at the top axis of each panel in Figure~\ref{4-plot}. For example, in the case of the $M_r<$ -20 sample, the physical range covered by our measurements is approximately $10 \hkpc < r < 300 \hkpc$, which is mainly probing the spatial distribution within halos.

%Wtheta results
The points in Figure~\ref{4-plot} show the $\wtheta$ measurements and the error bars are estimated from jackknife resampling, as described in \S\ref{measurement} (they are the diagonal values of the covariance matrix). The $M_r<$-18 sample is significantly noisier than the other three because it is the smallest of our galaxy samples (see Table~\ref{gal_summary}). The amplitude of $\wtheta$ is highest for the least luminous sample and drops progressively with luminosity. This is simply due to the fact that more luminous samples extend further in redshift, resulting in more uncorrelated galaxy pairs in each angular bin that dilute the clustering signal. 

As we discussed in the previous section, we expect fiber collision errors to be small in these measurements. However, there are still some galaxy pairs lost and gained in special cases where the nearest neighbor correction applied to collision triplets and higher multiplicity collision groups causes galaxies to incorrectly enter or exit the volume-limited sample. One of the advantages of using the angular correlation function is that errors due to fiber collisions should appear as a sharp feature at $55\arcsec$. An inspection of Figure~\ref{4-plot} shows no such significant features, except perhaps for a small feature in the case of the $M_r<$-19 sample. The $M_r<$-18 sample measurement shows two small discontinuities at small scales, but these occur between bins four and five and again between bins seven and eight, whereas the fiber collision scale occurs between bins eight and nine. We think that it is more likely that these small scale discontinuities are due to noise, given that they are similar in amplitude to the size of the data error bars, and that they occur at the wrong scales to be obviously caused by fiber collisions. We therefore conclude that fiber collision errors are indeed likely small, as expected. However, we emphasize that our analysis method has not eliminated fiber collision incompleteness and that it is definitely present in our measurements, as we discuss in detail in the Appendix.

All four correlation functions look approximately like power laws by eye and the most luminous sample appears to have a somewhat steeper correlation function than that of the lower luminosity samples. Before we can fit any model to our measurements we must first estimate covariance matrices. We do this using jackknife resampling, as described in \S\ref{measurement}. Figure~\ref{covar} shows the correlation matrix, which is the covariance matrix normalized by its diagonal elements, in the case of the $M_r<$-18 sample. The matrix clearly shows that nearby angular bins are highly correlated with each other, especially at large angular scales.

%We quantify this next by fitting a power law model to our measurements

\subsection{Power Law Fitting}
\label{power_law}

%Why a power law
Galaxy correlation functions are approximately shaped like power laws and so the power law model is often used to quantify their shape and amplitude. However, the near power-law shape of the galaxy correlation function is largely a coincidence \citep{Watson2011}, and it has been shown that the correlation function is not well described by a power law in a statistical sense, especially at the scales corresponding to the size of the typical dark matter halos that contain bright galaxies \citep{Zehavi2004}. Power law models are thus not accurate models and they do not directly yield a physical understanding of galaxy clustering. However, they are useful as a descriptive tool for quantifying the overall slope of the correlation function and for comparing the slopes of different galaxy samples. The inner slope of the density profile of satellite galaxies in halos directly affects the slope of the 3D correlation function on small scales, which in turn directly affects the slope of the angular correlation function. A steeper density profile for satellite galaxies should translate into a steeper $\wtheta$ \citep[e.g.][]{Peebles1980,Efstathiou1991,Watson2010}.  

\begin{figure}
\centering
\includegraphics[width=3.5in,height=3.5in]{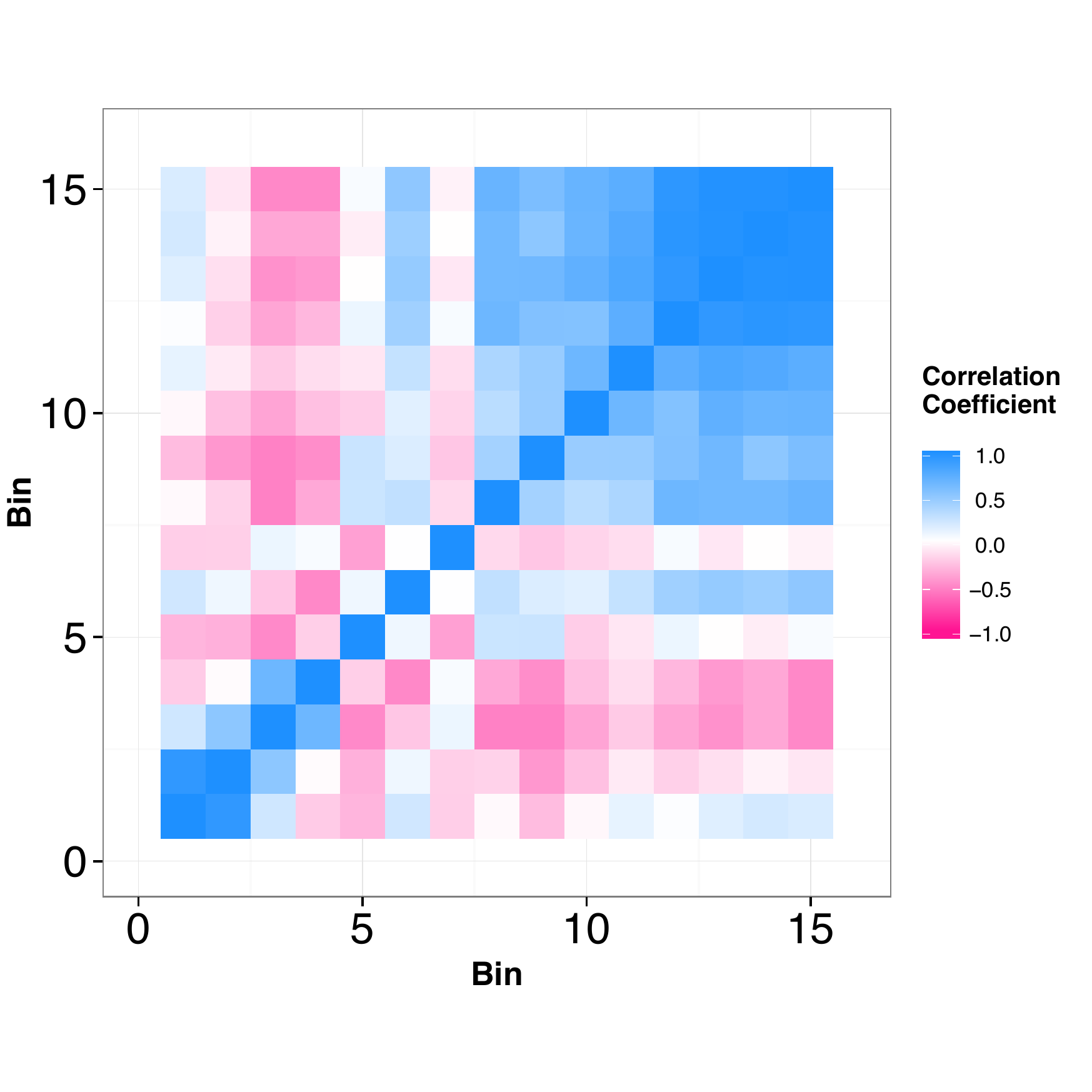} 
\vspace{-1cm}
\caption{Correlation matrix for the angular correlation function of the $M_r<$-18 sample, derived from 100 jackknife resamplings of the data on the sky. The correlation matrix is simply the covariance matrix normalized by its diagonal elements and we compute it as described in \S\ref{measurement}.
}
\label{covar}
\end{figure}

%fitting
We fit a power law model to our $\wtheta$ measurements using the MCMC code \texttt{emcee} \citep{emcee}, which we describe in more detail in \S\ref{spatialdistribution}. We calculate the $\chi^2$ value for each model parameter combination using
\begin{equation}
\chi^2 =\sum_{ij} (\omega_i -\omega_{\mathrm{model},i}) C^{-1}_{ij}(\omega_j -\omega_{\mathrm{model},j}),
\end{equation}
where $\omega_i$ and $\omega_{\mathrm{model},i}$ are the data and model correlation function in bin $i$, and $C^{-1}_{ij}$ is the inverse of the jackknife covariance matrix from Equation~\ref{covar_jack}. 

\begin{figure}
\centering
\includegraphics[width=3.5in,height=3in]{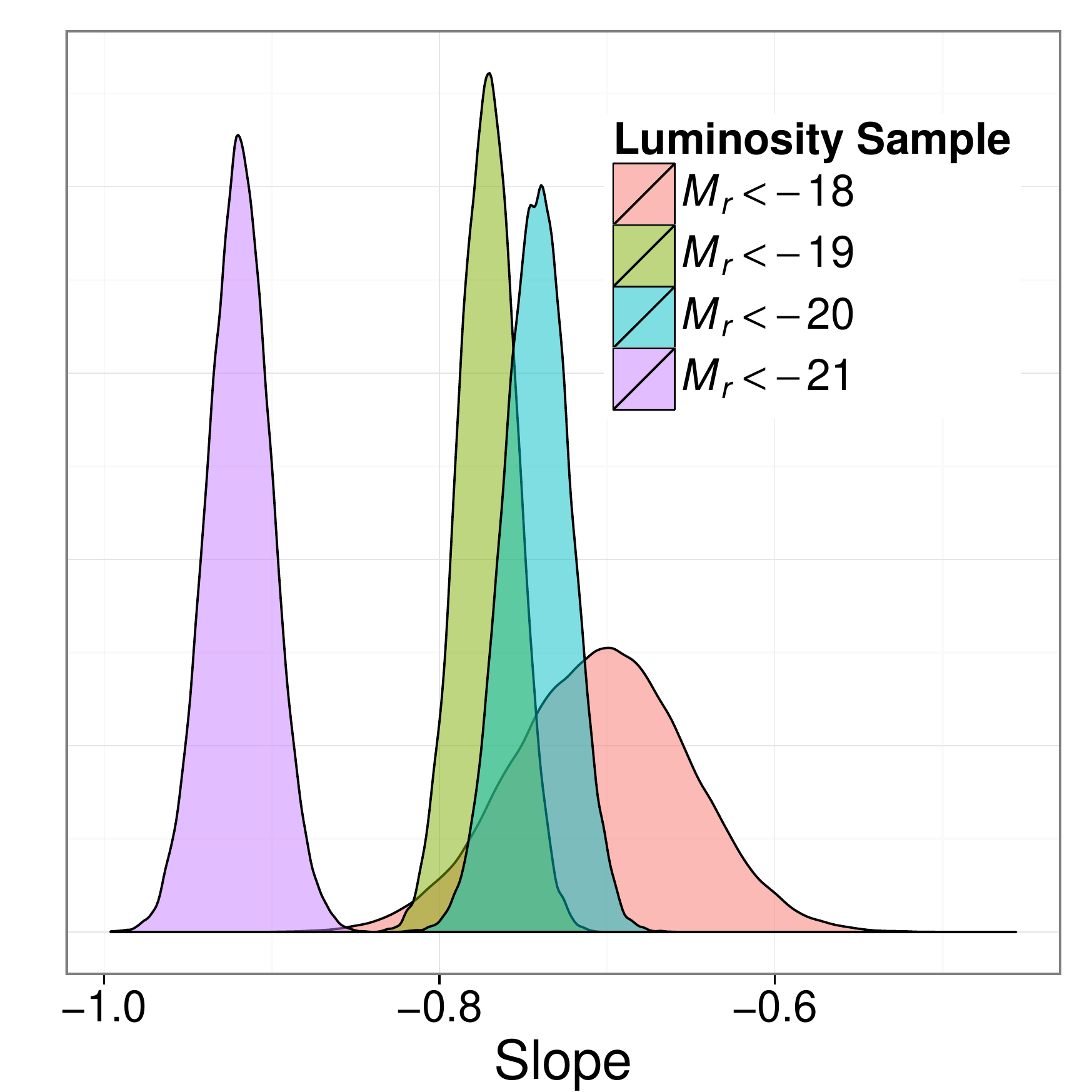} 
\caption{Probability density functions of slope for the power-law fits to the angular correlation functions of our four luminosity samples.  The clustering of the most luminous galaxies exhibits a significantly steeper slope than that of the less luminous galaxies.}
\label{contour}
\vspace{0.2cm}
\end{figure}

%fit results
Figure \ref{4-plot} shows a random sampling of power laws drawn from the best-fitting 68\% of models in the MCMC chains. We list the median values and 68\% confidence intervals of power-law slopes for all four samples in Table~\ref{gal_summary}, as well as the corresponding values of $\chi^2$ per degrees of freedom. The correlation matrix for the $M_r<$-18 sample is shown in Figure \ref{covar}. Finally, in Figure~\ref{contour}, we show the posterior probability density function of slope for each luminosity sample, as given from the MCMC chains. The best-fit $\chi^2$ values indicate that a power law functional form provides a good statistical description of the shape of $\wtheta$ for all four luminosity samples (the $M_r<$ -18 sample has a $p$-value of 0.12). Furthermore, the fit results show that the most luminous galaxy sample, $M_r<$ -21, has a significantly steeper power-law slope than the less luminous samples, while there is no trend in the steepness of the slope for the lower luminosity samples. This result seems to confirm the results of \cite{Watson2012}, who found that only luminous galaxies had steep satellite density profiles. In the next section we fit our clustering measurements with a halo model in order to directly probe what constraints we can place on the distribution of satellite galaxies within halos.

%-------------------------------
%-------------------------------

%================================================
\section{Halo Modeling}
\label{model}

%numerical modeling
Most previous studies fitting halo models to clustering measurements, including \cite{Watson2010,Watson2012}, have used an analytic framework to compute the correlation function. This framework requires analytic approximations for the halo mass function, the large scale bias of halos, and the halo density profile and it combines them together with a parameterized HOD to predict the distribution of galaxy pairs. Analytic halo models are fast and reasonably accurate; however, one should be cautious before trusting them at better than the $\sim10-20\%$ level. In this paper, we adopt a fully numerical procedure that eliminates most of the systematic errors that are present in analytic models.  We populate dark matter halos in cosmological N-body simulations with galaxies according to our adopted HOD, we then construct mock galaxy samples with similar selection as our SDSS samples, and we measure $\wtheta$ from the mocks in the same way as we do from the SDSS data. A few recent studies have used similar numerical modeling to fit galaxy clustering measurements \citep{White2011,Parejko2013,Reid2014}.

\subsection{Simulations and Halo Catalogues}

%lasdamas
We obtain halo catalogues from the Large Suite of Dark Matter Simulations \citep[LasDamas;][]{lasdamas} project\footnote{http://lss.phy.vanderbilt.edu/lasdamas/overview.html}. LasDamas  consists of many realizations of dark matter N-body simulations for a few different box size formats. The goal of the project is to create a large number of realistic mock galaxy catalogs for several luminosity samples in the SDSS in order to assist in the modeling of galaxy clustering measurements. For each luminosity sample that we model, we use a set of LasDamas simulations with appropriate box size and mass resolution, which are listed in Table~\ref{simtable}. All the simulations adopt a $\Lambda$CDM cosmological model with parameter values that are consistent with WMAP5 \citep{Dunkley2009}. The particle distributions were evolved using the code \texttt{GADGET-2} \citep{Springel2005}. Halos were identified from the dark matter distributions using a friends-of-friends (FoF) algorithm with a linking length of 0.2 times the mean inter-particle separation. We use halo catalogues from ten independent realizations (seeded with the same primordial power spectrum, but different random phases) when we model our clustering measurements in order to address cosmic variance errors in our analysis. We discuss this further in \S\ref{fittingroutine}.

%-------------------------------
\begin{deluxetable}{lccccc}
\tablecolumns{6}
\tabletypesize{\scriptsize}
\tablewidth{0pt}
\tablecaption{LasDamas Simulation Properties}
\tablehead{
Name & Sample & $L_\mathrm{box}$ & $N_{\mathrm{part}}$ & $m_{\mathrm{part}}$ & $r_\mathrm{soft}$ \\ 
& & (Mpc/$h$) & & ($10^{10} M_\odot/h$) & (kpc/$h$) \vspace{-0.15cm}\\ 
}
\startdata
\noindent
\vspace{0.2cm}
Consuelo & -18, -19 &420 & $1400^3$ & 0.187 & 8 \\
\vspace{0.2cm}
Esmeralda & -20 & 640 & $1250^3$ & 0.931& 15 \\
\vspace{0.2cm}
Carmen &  -21 & 1000 & $1120^3$ & 4.938& 25 \vspace{-0.25cm}\\
\enddata
\tablecomments{For each LasDamas simulation, the table lists the absolute magnitude limit for the galaxy sample modeled, the simulation box size, the number of particles, the particle mass, and the force softening scale. Ten realizations of each box were used in the analysis.}
\label{simtable}
\end{deluxetable}
%-------------------------------

\subsection{HOD Formalism}
\label{HOD_modelling_section}

%P(N|M)
We use the halo occupation distribution \citep[HOD; e.g.,][]{Berlind2002HOD} framework to create mock galaxy distributions from the dark matter halo catalogues. The HOD completely describes the bias between galaxies and dark matter by specifying the number and spatial positions of galaxies within halos. We first parameterize the probability distribution $P(N|M)$ that a dark matter halo of mass $M$ contains $N$ galaxies. We adopt the specific formalism introduced by \cite{ZhengCoilZehavi}, which separates central and satellite galaxies as motivated by theoretical results \citep{Kravtsov2004,Zheng2005}. The average number of central galaxies as a function of halo mass is essentially a smooth step function that rises from zero to one,
\begin{equation}
\label{hod1}
\langle N_\mathrm{cen}(M) \rangle = \frac{1}{2}\left [ 1 + \mathrm{erf}\left(\frac{\log M-\log \Mmin}{ \sigma_{\log M}} \right) \right],
 \end{equation}
where $\Mmin$ is the mass at which half the halos contain a central galaxy, and $\sigma_{\log M}$ controls the smoothness of the cutoff. The form of this function comes from the assumption that the scatter in the halo mass vs. galaxy luminosity relation has a lognormal form. The average number of satellite galaxies as a function of halo mass is essentially a power law with the same cutoff applied,
 \begin{equation}
 \label{hod2}
\langle N_\mathrm{sat}(M) \rangle = \langle N_\mathrm{cen}(M) \rangle \left(\frac{M-M_0}{M_1}\right)^\alpha,
 \end{equation}
where $M_0$ is the halo mass below which there are no satellite galaxies, $M_1$ is the halo mass that contains on average one satellite galaxy \footnote{This is not exactly true unless $\langle N_\mathrm{cen}\rangle=1$ and $M_1$$\gg$$M_0$. However, for the samples we consider here, this is close to correct.}, and $\alpha$ is the slope of the power law relation. Once we have specified the mean number of centrals in a halo using Equation~\ref{hod1}, we place an actual central galaxy in that halo using a probability equal to $\langle N_\mathrm{cen}\rangle$ (e.g., if $\langle N_\mathrm{cen}\rangle=0.7$, we give the halo a 70\% chance of actually containing a central galaxy). Likewise, once we have specified the mean number of satellites in a halo using Equation~\ref{hod2}, we draw an actual number of satellites for that halo from a Poisson distribution.

\subsection{Spatial Distribution of Galaxies Within Halos}
\label{spatialdistribution}

%spatial distribution within halos
Once we know how many galaxies a halo receives we have to decide where to put them. We place each central galaxy at the deepest location of its halo's gravitational potential well, which we calculate from the dark matter particles in the halo. For satellite galaxies, we adopt the methodology of \cite{Watson2010} and introduce a Generalized Navarro-Frenk-White (GNFW) density profile
\begin{equation}
\rho_{gal}(r) = \frac{\rho_s}{(\frac{r}{r_s})^\gamma (1 + \frac{r}{r_s})^{3-\gamma}},
\label{gamma}
\end{equation}
where the slope of the profile transitions from -$\gamma$ in the inner regions of the halo to -3 in the outer regions. As with a NFW profile, the transition scale 
depends on the concentration, but we allow the concentration of satellite galaxies to differ from that of dark matter through the parameter $\fgal$ 
\begin{equation}
\label{fgal}
c_\mathrm{gal} = \fgal \times c_\mathrm{DM}.
\end{equation}
For the dark matter concentration we adopt the modified \cite{Bullock2001} relation from \citet{ZhengCoilZehavi}
\begin{equation}
c_\mathrm{DM} = 11\left( \frac{M}{M_\star} \right) ^{-0.13}.
\end{equation}
The GNFW profile thus has two free parameters, $\gamma$, and $\fgal$, and we draw random radii from this profile to determine the positions of satellite galaxies within each halo. Note that values of $\gamma = \fgal =1$ recover an NFW profile. Drawing satellite positions from an analytic profile instead of using actual particle positions allows us to avoid force resolution errors that occur at the smallest scales we consider. Models of this type have been used to model the inner slope of the dark matter density profile \citep[e.g.,][]{Fukushige2004,Reed2005}. 

%-------------------------------
\begin{figure}
\centering
\includegraphics[width=3.5in,height=3.5in]{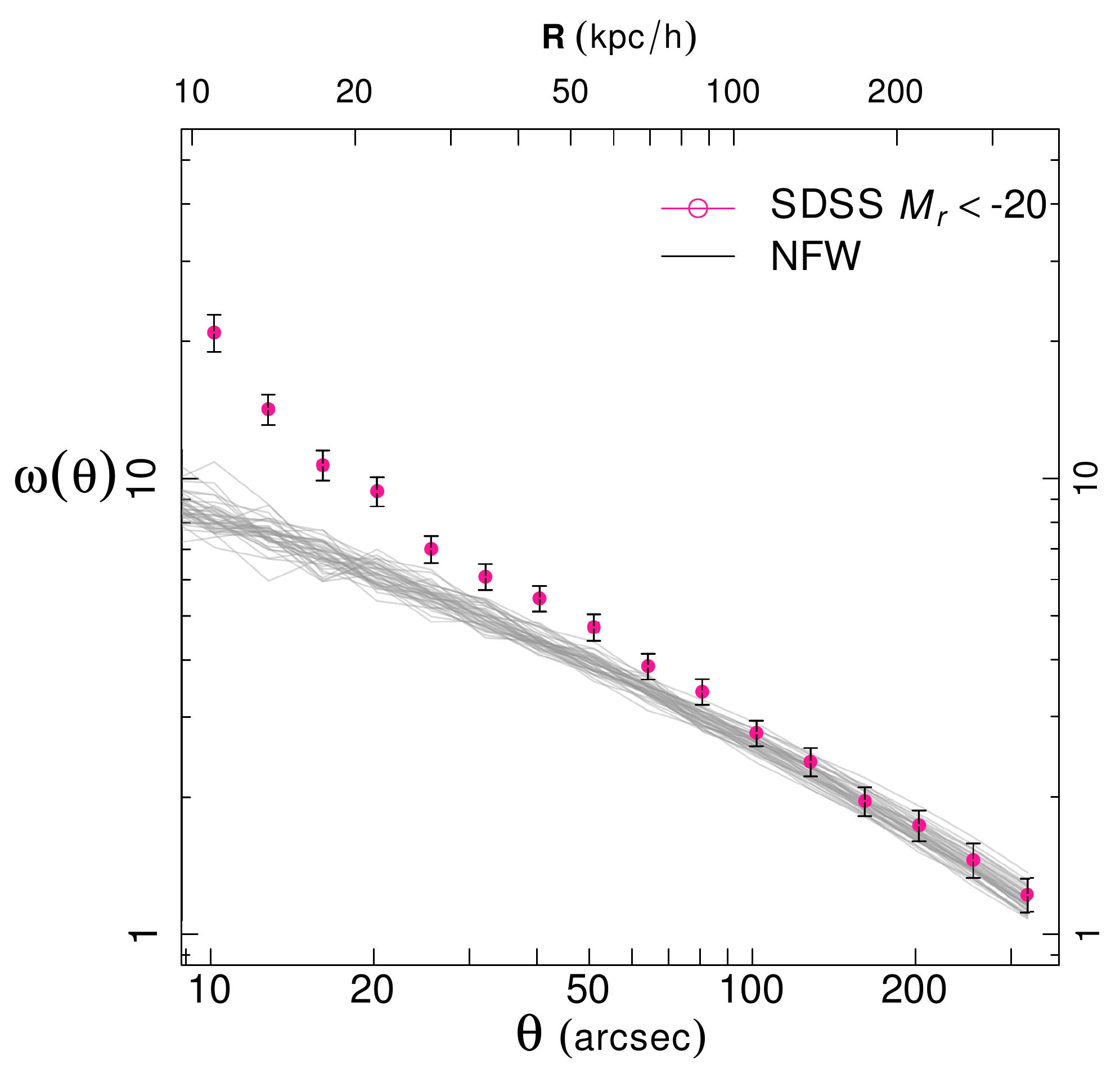} 
\caption{
Angular correlation function for SDSS galaxies with $M_r<$ -20 compared to a model where satellite galaxies within dark matter halos follow a NFW density profile. Points show $\wtheta$ for SDSS galaxies (also shown in Figure~\ref{4-plot}). Curves show measurement from several independent mock galaxy catalogs that are constructed by populating dark matter halos in N-body simulations with galaxies. Each sufficiently massive halo gets a central galaxy that is placed at the halo center, while any additional satellite galaxies are distributed according to an NFW density profile. The bottom axis shows the angular scale in units of arcsec and the top axis shows the corresponding physical scale at the median redshift of the $M_r<$ -20 sample.
}
\label{wtheta_nfw}
\vspace{0.5cm}
\end{figure}
%-------------------------------

%demonstration that NFW does not work
Before exploring the parameter space of our flexible GNFW model, we briefly investigate whether the NFW model can reproduce our $\wtheta$ measurements. We construct mock catalogs for the $M_r<$ -20 SDSS sample using halo catalogs from the Esmeralda simulations, as detailed above. For this test we adopt values for the HOD parameters outlined in \S\ref{HOD_modelling_section} that have been found to yield a projected correlation function $w_p(r_p)$ that agrees with $M_r<$ -20 SDSS galaxies on scales larger than $100\hkpc$ (McBride, private communication). We then adopt $\gamma = \fgal = 1$ for our satellite profile, which corresponds to a NFW profile. Figure~\ref{wtheta_nfw} shows $\wtheta$ for several independent mock catalogs compared to our SDSS measurements. The NFW mock catalogs go from faithfully reproducing the clustering at high angular separations to under-predicting the observed clustering on the very small scales. We therefore see the same tension as \cite{Masjedi2006} and this further motivates us to explore alternative density profiles for satellite galaxies.

\subsection{Computing $\wtheta$}
\label{computing_wtheta}

%spherical geometry
Once we have populated a N-body simulation with galaxies as outlined above, we place the observer at the center of the box, compute spherical coordinates, and throw out galaxies that lie outside the redshift limits of the sample we are trying to model. We do not include redshift space distortions in our analysis since they do not affect the angular clustering. Each resulting mock catalog covers the full celestial sphere and thus contains about 12 times more volume than the corresponding SDSS sample. This guarantees that the cosmic variance and shot noise in the mock catalog are much lower than in the SDSS and will therefore not significantly degrade the precision of our results.

%estimator
We compute $\wtheta$ using the natural estimator,
\begin{equation}
\label{nestimator}
w(\theta) = \frac{DD}{RR} - 1.
\end{equation}
We choose this estimator because it does not include a $DR$ term, which is computationally much more expensive than $DD$ since the random catalog is much larger than the data catalog. The $RR$ term only needs to be computed once so when we perform our model parameter search we only have to compute $DD$ for each set of model parameter values. This estimator is different from the one shown in Equation~\ref{lsestimator}; however, on small scales and for a full sky geometry, these estimators yield indistinguishable results \citep{Kerscher2000}.

\subsection{Model Fitting}
\label{fittingroutine}

%-------------------------------
\begin{figure*}
\centering
\includegraphics[width=5in,height=5in]{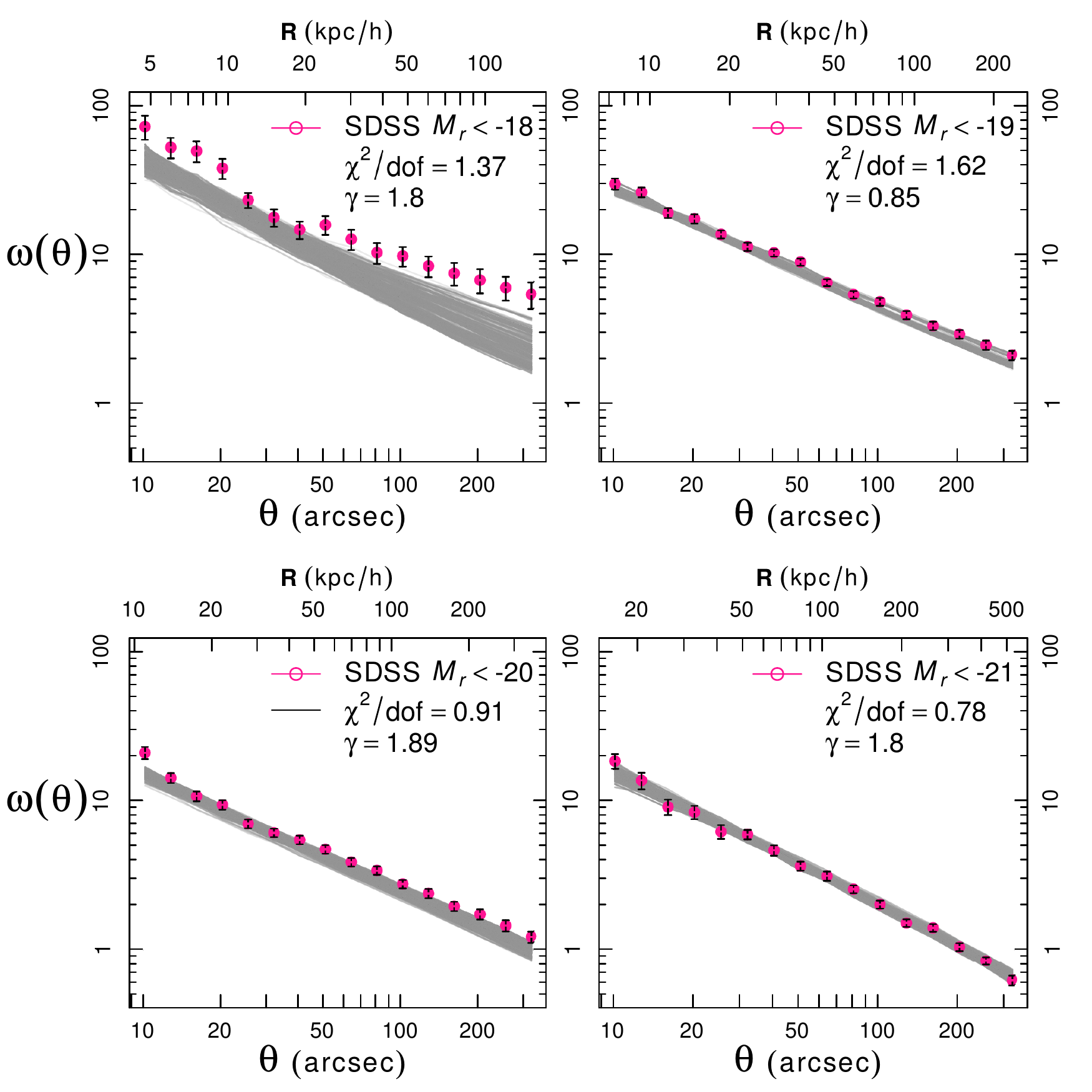} 
\caption{The angular correlation function of SDSS galaxies in four volume-limited samples, along with their halo model fits. The data measurements and overall layout are the same as in Fig.~\ref{4-plot}.  The gray lines show a selection of model correlation functions that are randomly drawn from the best-fitting 68\% of halo models in the MCMC chains. The median value of the satellite galaxy density profile inner slope $\gamma$, and the goodness of fit are listed in each panel.}
\label{HODfourplot}
\end{figure*}
%-------------------------------
%-------------------------------
\begin{deluxetable*}{lcccccccc}
\tablecolumns{9} 
\tabletypesize{\scriptsize}
\tablewidth{0pt}
\tablecaption{Median Values of Halo Model Parameters from MCMC Chains}
\tablehead{$M_r^\mathrm{lim}$ & $\sigma_{\log M}$ & $\log M_0$ & $\log {M_1}$ & $\alpha$ & $\gamma$ & $\fgal$ & $\nicefrac{\mathlarger{\chi}^2}{\mathrm{dof}}$ & $\nicefrac{P}{P_\mathrm{NFW}}$ \\}
\startdata
\vspace{0.2cm}
-18 & $0.48\hspace{0.1cm}_{-0.33}^{+0.36}$ & $8.14\hspace{0.1cm}_{-2.15}^{+2.34}$ & $12.87\hspace{0.1cm} _{-0.10}^{+0.16}$ & $1.10\hspace{0.1cm} _{-0.08}^{+0.10}$ & $1.80\hspace{0.1cm} _{-0.56}^{+0.46}$ & $1.11\hspace{0.1cm} _{-0.74}^{+0.62}$ & $\nicefrac{13.7}{10}$ & 3.13 \\
\vspace{0.2cm}
-19 & $0.14\hspace{0.1cm} _{-0.11}^{+0.19}$ & $9.30\hspace{0.1cm} _{-0.89}^{+1.01}$ & $12.90\hspace{0.1cm} _{-0.03}^{+0.04}$ & $1.11\hspace{0.1cm} _{-0.03}^{+0.03}$ & $0.85\hspace{0.1cm} _{-0.47}^{+0.61}$ & $1.33\hspace{0.1cm} _{-0.64}^{+0.46}$ &$\nicefrac{16.2}{10}$ & 1.17 \\
\vspace{0.2cm}
-20 & $0.49\hspace{0.1cm} _{-0.35}^{+0.34}$ & $10.10\hspace{0.1cm} _{-1.46}^{+1.59}$ & $13.44\hspace{0.1cm} _{-0.12}^{+0.14}$ & $1.33\hspace{0.1cm} _{-0.12}^{+0.15}$ & $1.89\hspace{0.1cm} _{-0.46}^{+0.27}$ & $0.51\hspace{0.1cm} _{-0.37}^{+0.80}$ & $\nicefrac{9.1}{10}$ & 6.33 \\

-21 & $0.54\hspace{0.1cm} _{-0.35}^{+0.31}$ & $10.72\hspace{0.1cm} _{-1.90}^{+1.80}$ & $14.03\hspace{0.1cm} _{-0.09}^{+0.10}$ & $1.63\hspace{0.1cm} _{-0.19}^{+0.22}$ & $1.80\hspace{0.1cm} _{-0.40}^{+0.24}$  & $0.99\hspace{0.1cm} _{-0.64}^{+0.65}$ & $\nicefrac{7.8}{10}$ & 9.32 \\
\enddata
\tablecomments{The median halo model parameter values, along with the middle 68\% interval, as measured from the MCMC chains. Also shown are the best-fit value of $\chi^2$, as well as $P/P_\mathrm{NFW}$, the ratio of probability between the median value of $\gamma$ and $\gamma=1$.}
\label{tableparams}
\end{deluxetable*}
%-------------------------------

%free parameters
We are most interested in constraining the inner slope of the satellite galaxy density profile, which is described by the $\gamma$ parameter. Even though this parameter plays a primary role in setting the shape of the correlation function on the small scales we are examining, it is not easy to disentangle its effect on $\wtheta$ from that of the other HOD parameters. We therefore allow all the following parameters to be free during our parameter search:
\begin{enumerate}\itemsep1pt
\item $\sigma_{\log M}$: Amount of scatter in the luminosity-mass relation for central galaxies.
\item $M_0$:  Mass below which halos contain no satellite galaxies.
\item $M_1$:  Mass at which halos contain on average one satellite galaxy.
\item $\alpha$: Slope of the power-law relation between the mean number of satellite galaxies and halo mass.
\item $\gamma$:  Inner slope of the number density profile for satellite galaxies within their halo.
\item $\fgal$: Concentration of satellite galaxies with respect to the dark matter concentration.
\end{enumerate} 
For each combination of the above six free parameters, we set $\Mmin$ to the value that recovers a total galaxy number density equal to that observed by the SDSS. 

%MCMC procedure
We perform a parameter search using the MCMC \texttt{emcee} code and algorithm described by \cite{emcee}. The algorithm is based on the affine invariant sampling algorithm proposed by \cite{Goodman2010}. It is fast, efficient, and easily parallelized. We use 500 ``walkers" to explore the parameter space in parallel. The basic procedure we follow each time we test a new location in our six dimensional parameter space is as follows. We first select a random halo catalog from among ten independent N-body realizations. This builds cosmic variance errors in our theoretical calculations directly into the modeling. We then use the halo catalog to determine the value of $\Mmin$ required to create a galaxy catalog with the observed mean number density. For each halo in the catalog, we use Equation~\ref{hod1} to determine whether the halo contains a central galaxy, and Equation~\ref{hod2} together with a Poisson distribution to choose the number of satellite galaxies. We then randomly draw satellite positions from the density profile shown in Equations~\ref{gamma} and~\ref{fgal}. We make an all-sky galaxy mock catalog and compute $\wtheta$ as described in \S\ref{computing_wtheta}. Finally, we estimate $\chi^2$ for the parameter combination using the jackknife covariance matrix described in \S\ref{measurement}. All six of our free parameters are given physically motived flat priors.  In particular, the satellite profile parameters $\gamma$ and $\fgal$ are allowed to vary over a broad range that includes the NFW profile. For all four luminosity samples, we find that we need approximately one million parameter combinations in order to get MCMC chains that converge.

\subsection{Halo Modeling Results}
\label{results}

%wtheta modeling results
Figure~\ref{HODfourplot} shows the resulting $\wtheta$ of our halo modeling in each luminosity bin. SDSS measurements and the overall figure layout are the same as in Figure~\ref{4-plot}, while the gray lines show a random sampling of halo model correlation functions drawn from the best-fitting 68\% of models in the MCMC chains. The lines thus illustrate the spread in $\wtheta$ for models that are consistent with the SDSS data. Each panel shows the median value of $\gamma$, as well as the value of $\nicefrac{\chi^2}{\mathrm{dof}}$ for the best fit model. With 16 angular bins and 6 free parameters, the number of degrees of freedom is equal to 10. Our halo model produces good fits to the angular clustering of all four luminosity samples. There is a slight tension in the case of the $M_r<$ -19 sample, but the difference between the model and the SDSS data is not statistically significant (the $p$-value for this sample is 0.094). We list the $\nicefrac{\chi^2}{\mathrm{dof}}$ values for all four samples in Table~\ref{tableparams}. Though it looks like the model is not a good fit to the data in the case of the $M_r<$-18 sample, we note that it is very misleading to perform Ò$\chi$ by eyeÓ because neighboring bins in $\wtheta$ are extremely correlated with each other, as shown in Figure~\ref{covar}.

%-------------------------------
\begin{figure}
\centering
\includegraphics[width=3in,height=3in]{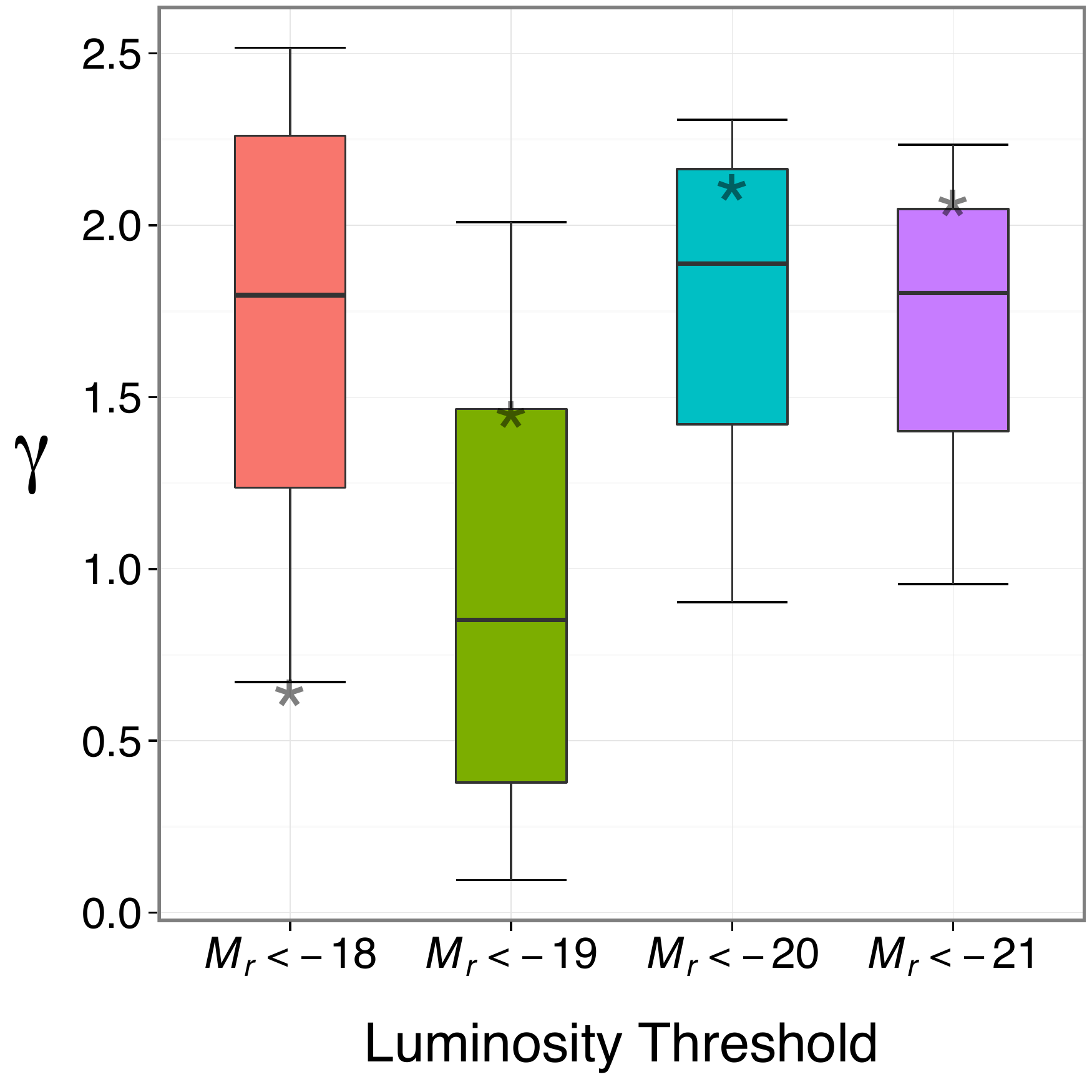} 
\caption{Luminosity dependence of the satellite galaxy density profile inner slope.  Each box and associated whiskers corresponds to a particular luminosity sample, as shown on the x-axis. The middle line in each box shows the median value of $\gamma$ from the MCMC chain, the vertical range of the box shows the middle $68\%$ of $\gamma$ values, and the whiskers extend to the middle $95\%$ of values. For the two high luminosity samples, a value of $\gamma = 1$, corresponding to the NFW density profile, is inconsistent with the SDSS data at approximately the $2\sigma$ level. Lower luminosity galaxies do not show this tension. The median values of $\gamma$ from \cite{Watson2012} are marked as asterisks.}
\label{gamma_fits}
\end{figure}
\vspace{0.2cm}
%-------------------------------

%gamma results
The main focus of this analysis is the inner slope $\gamma$ of the satellite galaxy density profile within halos. Figure~\ref{gamma_fits} shows the marginal distribution of $\gamma$ values from the MCMC chains as a function of galaxy luminosity. Specifically, the middle line in each box shows the median value of $\gamma$, the shaded box shows the 68\% confidence interval, and the whiskers show the 95\% confidence interval for $\gamma$. The two most luminous samples, $M_r<$ -20 and -21, both prefer fairly steep density profiles and are inconsistent with the NFW profile at approximately the $2\sigma$ level. Specifically, the fraction of points in the MCMC chain that have $\gamma>1$ is 97\% and 96\% for the $M_r<$ -21 and -20 samples, respectively. The less luminous $M_r<$ -19 sample prefers less steep profiles and is perfectly consistent with NFW. The lowest luminosity $M_r<$ -18 sample seems to favor steep profiles, but it has a broad $\gamma$ distribution and is not significantly inconsistent with NFW. The poor constraints for the least luminous sample are due to the small size of this galaxy sample. 

%Figure~\ref{gamma_fits} does not suggests a luminosity trend, where luminous satellite galaxies have steeper density profiles than lower luminosity galaxies; however, this trend is not very significant. 

The constraints on $\gamma$ are consistent with those from \cite{Watson2012}, denoted by the asterisks in Figure~\ref{gamma_fits}, at the $1\sigma$ level for the $M_r<$ -19, -20, and -21 samples. The $M_r<$-18 sample is consistent at approximately the $2\sigma$ level. The constraints are somewhat weaker in this paper because we reduce our sample size by only considering galaxies in plate overlap regions. Additionally, fiber collisions are a source of systematic error that was not included explicitly in our model. Though we do not expect this error to be large, it might be affecting the $M_r<$-18 measurement in a subtle way that is contributing to the tension with \cite{Watson2012}. We list the median and 68\% confidence intervals for all four samples in Table~\ref{tableparams}. We note that a cursory examination of Figure~\ref{wtheta_nfw} may lead to the impression that the NFW model is ruled out at higher significance than $2\sigma$. However the models shown in Figure~\ref{wtheta_nfw} were fit to $w_p(r_p)$ on larger scales, not $\wtheta$ on very small scales. Moreover, as stated previously, it is misleading to perform Ò$\chi$ by eye due to the high degree of correlation between bins.

%difference from NFW
We now quantify the extent to which the $\gamma$ values preferred by our clustering measurements are more likely than the $\gamma=1$ NFW case. We calculate the number of accepted parameter combinations in the MCMC chain that have values of $\gamma$ in a bin of width $\pm 0.1$ that is centered around the median value of $\gamma$. We then do the same for a bin centered around $\gamma=1$ and take the ratio of these two numbers, which we call $\nicefrac{P}{P_\mathrm{NFW}}$. This yields the relative likelihood of the two $\gamma$ values given the measured correlation function. We find that the steep slopes measured for the $M_r<$ -20 and -21 samples are 6.3 and 9.3 times more likely than $\gamma=1$, while the slopes measured for the less luminous samples are only 3.1 and 1.2 times more likely than $\gamma=1$. We list these values in Table~\ref{tableparams}. 

%-------------------------------
\begin{figure}
\centering
\includegraphics[width=3in,height=3in]{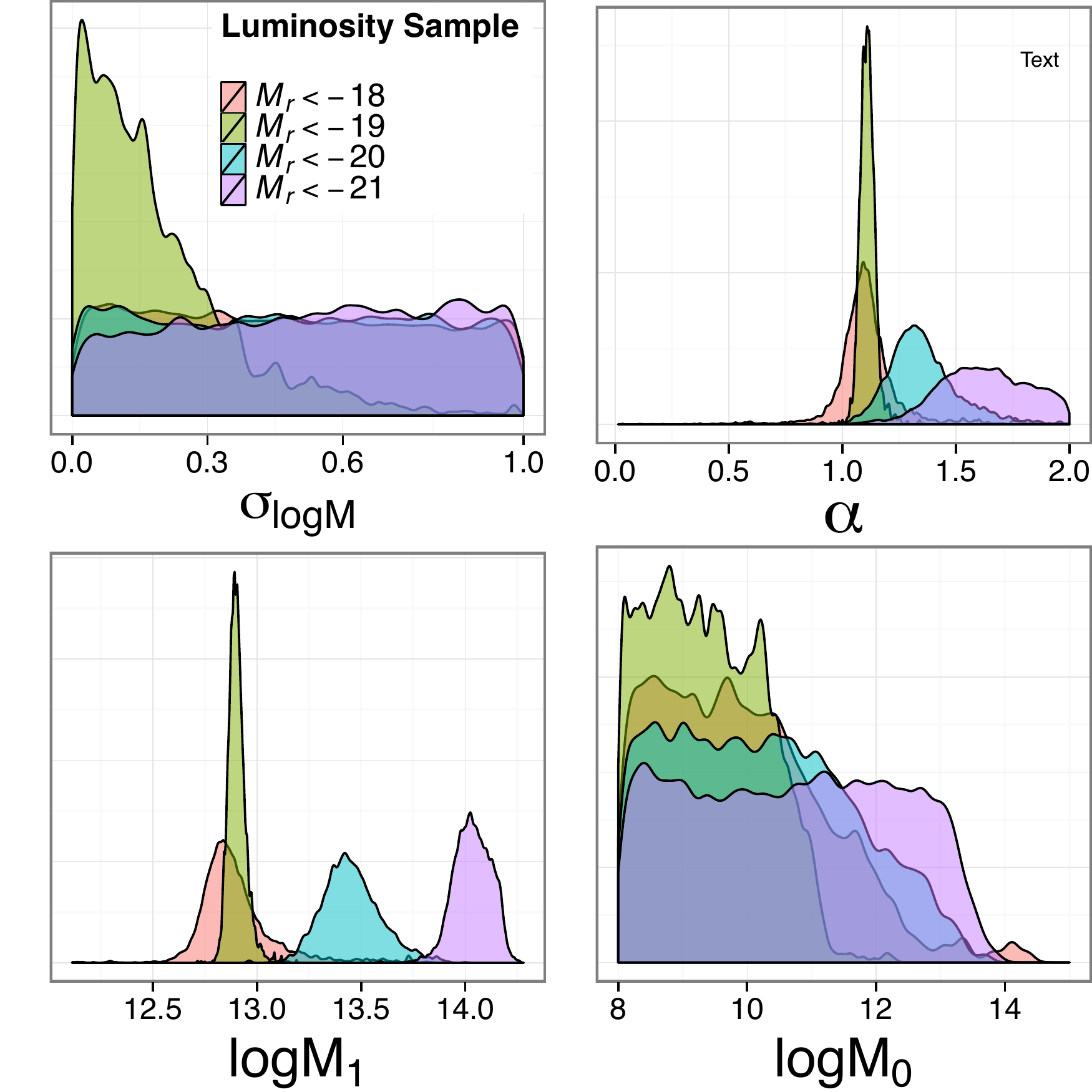} 
\caption{The probability distributions for the four HOD parameters that control the number of galaxies a dark matter halo of a given mass will receive. Each panel shows a different parameter, while the four distributions in each panel show results for our four luminosity samples. The parameters $\sigma_{\log M}$ and $M_0$ are largely unconstrained by our measurements and the bounds of their distributions reflect their prior flat distributions. $\sigma_{\log M}$ was restricted to the range $0-1$, while $M_0$ was restricted to values greater than $10^8\msun$.}
\label{HOD_fits}
\vspace{0.5cm}
\end{figure}
%-------------------------------

%other parameter results
Figure~\ref{HOD_fits} shows the final probability distributions of the HOD parameters that determine the number of galaxies as a function of halo mass. The parameter controlling the shape of the low mass cutoff for central galaxies $\sigma_{\log M}$ is poorly constrained by the angular correlation function on the small scales we consider in this study. This is due to the fact that on these scales most galaxy pairs come from within a single halo and so the low mass regime where a halo either has zero or one galaxy is not very important. The distributions of $\sigma_{\log M}$ are bound by values of 0 and 1 as this was the prior that we adopted for this parameter. The parameter $M_0$ is also very poorly constrained because it cuts off the satellite occupation number on a sufficiently small mass scale where the expected number of satellites is already much less than one. This result is consistent with other studies \citep[e.g.,][]{Zehavi2011}. The lower bound for $M_0$ at $10^8\msun$ is once again due to the prior we adopted for this parameter. Though these priors are physically motivated, we have checked that relaxing them does not significantly change our conclusions about the slope $\gamma$. The two parameters that control the number of satellites a halo receives, $\alpha$ and $M_1$, are much better constrained.  More luminous samples have a higher mass $M_1$ at which they typically contain a single satellite and they have a steeper relation between the number of satellites and halo mass.  These trends are consistent with \cite{Zehavi2011} though the values we find are somewhat higher for both $M_1$ and $\alpha$. This could be due to the different information content of $\wtheta$ compared to $w_p(r_p)$, or it could be due the difference between our numerical modeling compared to the analytic halo model used in \cite{Zehavi2011}, or it could be due to the extra freedom we have added to the HOD model by using a GNFW profile. We note that for the purpose of this paper we are mainly interested in the $\gamma$ parameter and these parameters therefore act as nuisance parameters. We list the median and 68\% confidence intervals for all model parameters in Table~\ref{tableparams}. We also note that the galaxy concentration parameter $\fgal$ is very poorly constrained.

\subsection{Power Law vs. Halo Model}

%differences between power law vs. halo model results
Both the power-law fits and the halo model have suggested that luminous satellite galaxies have a steeper density profile than lower luminosity galaxies. However, the two analyses also show some differences. In the case of the power law slope, $M_r<$ -21 galaxies have a steeper $\wtheta$ than less luminous samples and this difference is highly significant. On the other hand, when using the halo model, $M_r<$ -20 galaxies have consistent values of $\gamma$ with $M_r<$ -21 galaxies and discrepancies with less luminous galaxies are less significant. We now take a closer look at these results to determine if they are consistent with each other.
 
%naive comparison
We first make a naive comparison between the power law slope and the value of $\gamma$. If we assume that the galaxy correlation function $\xi(r)$ is dominated by central-satellite pairs in halos of mass $\sim M_1$ that only have a single satellite galaxy, then its slope should essentially be the same as the slope of the satellite galaxy density profile within these halos. This is not a bad assumption since more massive halos that contain more than one satellite galaxy are relatively rare. Therefore, the slope of $\xi(r)$ should be equal to -$\gamma$. If we further assume that $\xi(r)$ is a perfect power law, then the angular correlation function $\wtheta$ should also be a power law with a slope that is shallower by +1 \citep{Totsuji1969}. Therefore, the slope of $\wtheta$ should be $1-\gamma$, or conversely, $\gamma$ should be $1-slope$. Using this simple transformation, we can check whether the power law slopes that we found in \S\ref{power_law} are consistent with the $\gamma$ distributions shown in Figure~\ref{gamma_fits}. The values of $\gamma$ inferred from the power law slopes of $\wtheta$ are 1.7, 1.77, 1.76, and 1.92 for the $M_r<$ -18, -19, -20, and -21 samples, respectively. These values are perfectly consistent with the broad distributions shown in Figure~\ref{gamma_fits}.

%fit power law to gamma mocks
We next perform a more sophisticated test to assess the relationship between the slope of the power law model and $\gamma$. We fit a power law to measurements of $\wtheta$ made from mock galaxy catalogues produced using the best-fit HOD model. We then compare the recovered power law slopes and compare them to the input values of $\gamma$. We find that for three out of four samples, $slope \sim 1-\gamma$, as expected.  For the $M_r<$ -19 sample, the power-law slope is somewhat steeper than $1-\gamma$. We conclude from these explorations that our results from fitting power laws and halo models are consistent with each other, and that constraints on $\gamma$ are weaker than on the power law slope due to marginalization over all the other HOD parameters.

%================================================
\section{Summary \& Discussion}
\label{discussion}

%summary of goal and measurements
The goal of this paper is to probe the radial number density profile of satellite galaxies within dark matter halos using SDSS clustering measurements. We wish to test the results of \cite{Watson2012}, who found that luminous satellite galaxies (SDSS, $M_r<$ -20) have significantly steeper density profiles than the NFW profile on scales smaller than $\lesssim 40\hkpc$. Unfortunately, clustering measurements on these scales are strongly affected by fiber collision incompleteness, making it important to verify this result with different measurements and modeling methodology. We used the angular correlation function $\wtheta$ as our clustering statistic of choice because it is fairly insensitive to fiber collision errors. Moreover, we restricted our measurements to plate overlap regions on the sky, where most fiber collided galaxy pairs are recovered because of repeat observations. We measured $\wtheta$ on four volume-limited samples with absolute $r$-band limits of $M_r<$ -18, -19, -20, and -21, on scales in the range $7-320\arcsec$. These angular scales correspond to physical scales that are within the virial radii of dark matter halos expected to host these galaxies, even at the far redshift of each sample. Our measurements thus directly probe the spatial distribution of galaxy pairs within halos.

%summary of results
Motivated by the approximately power-law shape of our correlation function measurements, we first fit a power law function to $\wtheta$ in order to quantify its slope. We found that the most luminous galaxies ($M_r<$ -21) have a significantly steeper correlation function than the lower luminosity samples. We then used the more physically motivated halo model to determine what constraints our $\wtheta$ measurements place on the density distribution of satellite galaxies within halos. We used a fully numerical modeling procedure that populates dark matter halos in N-body simulations with galaxies, creates mock SDSS samples, and computes $\wtheta$ the same way as it is done in the real galaxy data. This method is computationally expensive, but it minimizes systematic errors in the modeling. The key ingredient in our halo model is a generalized density profile for satellite galaxies, whose inner slope is a free parameter. After marginalizing over other parameters in our halo model, we found that the two more luminous galaxy samples ($M_r<$ -20 and -21) prefer a satellite density profile that is substantially steeper than the NFW profile. The NFW profile is discrepant at the $2\sigma$ level for these galaxy samples. We found that the lower luminosity samples do not constrain the satellite inner profile slope as well and they are consistent with NFW.

%comparison to Watson et al. 
Our results are qualitatively consistent with those of \cite{Watson2012} who also found that satellite galaxies more luminous than $M_r<$ -20 have steeper density profiles than NFW. Our results are also quantitatively consistent, as our marginal distributions of the inner profile slope overlap nicely. The main differences between our two studies are that (1) \cite{Watson2012} found somewhat shallower inner profiles than we did for the least luminous ($M_r<$ -18) galaxies, and (2) their constraints on the inner profile slope of the most luminous galaxies are tighter than ours. These differences allowed them to detect a significant luminosity trend in the spatial distribution of satellite galaxies, while we cannot do the same with confidence. The loss of statistical significance in our study is mainly due to the lower information content of $\wtheta$ compared to $w_p(r_p)$, as well as to the fact that we reduce our sample size by only considering galaxies in plate overlap regions. On the other hand, our constraints are less likely to be affected by errors in the halo modeling. In addition, the two studies are affected by fiber collision incompleteness in different ways. Overall, the agreement between the two studies despite the different measurement and modeling methods lends credibility to the main conclusion that the spatial distribution of luminous satellite galaxies is steeper than that of the underlying dark matter.

%does NFW accurately represent DM
Before making claims about how well satellite galaxies trace the dark matter distribution, we need to consider whether the NFW model is itself an accurate representation of the density profile of dark matter. Though the NFW profile has been shown to provide an imperfect description of the structure of dark matter halos in collisionless N-body simulations \citep[e.g.,][]{Navarro2004,Merritt2005,Gao2008,Navarro2010,Ludlow2013}, the departures shown by these studies are not large and the NFW model remains consistent with simulation results at the $\sim 10-20$\% level \citep{Benson2010}. However, it is far less safe an assumption that the density profiles of halos in collisionless simulations represent reality given that they completely ignore the effects of baryons. This is especially true for the very small scales we consider in this paper, since baryons dominate the mass budget at the centers of halos. Some theoretical studies argue that the condensation of baryons at the centers of halos should steepen the dark matter density profile \citep[e.g.,][]{Gnedin2011}, while others argue the opposite \citep[e.g.,][]{DelPopolo2012}. Observational studies using weak lensing measurements have found that the density profiles at of clusters are either consistent with or shallower than NFW \citep[e.g.,][]{Mandelbaum2008,Newman2013}, though the measurements are noisy on the small scales we care about here. 

%should satellite galaxies trace DM
It is not necessarily surprising that satellite galaxies do not trace the underlying mass distribution. Galaxies are extended massive objects and they should thus experience dynamical effects such as dynamical friction and tidal stripping of mass and stars, which do not affect dark matter particles. These mechanisms could act to steepen the density profile of satellites. Moreover, this could be a luminosity dependent process. To study these effects, it would be illuminating to compare our satellite profile results with the distribution of dark matter subhalos within host halos, since satellite galaxies presumably occupy these systems. However, this comparison will have to wait for  simulations of sufficient volume and particle mass resolution to be able to accurately measure the distribution of massive subhalos at scales of only a few tens of kpc from the center of host halos. It would be even better to compare our results with predictions from hydrodynamic simulations that include baryonic processes such as gas cooling and feedback, which can affect the density profiles of halos. Our measurements can help to constrain these processes. Simulations that have both sufficient volume and resolution to make such predictions are now becoming possible. For example, the Illustris simulation has already enabled a prediction of the density profile of luminous satellite galaxies on small scales \citep{Genel2014}. Though this result is a bit too noisy to be tested against our measurements, the next generation of simulations should be more than adequate for making this comparison.

\acknowledgments
We sincerely thank Dan Foreman-Mackey, Ryan Scranton, and Manodeep Sinha for valuable discussions and software. We thank the anonymous referee for helping us improve the clarity of the paper and for suggesting the fiber collision test outlined in the Appendix. J.A.P. and A.A.B. were supported by the National Science Foundation (NSF) through NSF grant AST-1109789. The simulations used in this paper were produced by the LasDamas project (http://lss.phy.vanderbilt.edu/lasdamas/); we thank NSF XSEDE for providing the computational resources for LasDamas. Some of the computational facilities used in this project were provided by the Vanderbilt Advanced Computing Center for Research and Education (ACCRE).

%================================================
\appendix

\section{Fiber Collision Incompleteness}

As we discussed in \S\ref{Data_Sample}, our method of reducing the effects of fiber collisions consists of three parts. First, we restrict our samples to the plate overlap regions, where the vast majority of collided galaxies are recovered. Restricting our samples to the plate overlap regions guarantees that all cases of collision pairs are recovered. These are cases where there are only two galaxies within 55$\arcsec$ of each other. Additionally, some cases of collision triplets are recovered. These are cases where one galaxy is within 55$\arcsec$ of two other galaxies, but these others are not within 55$\arcsec$ of each other. However, other cases of collision triplets are not recovered, such as when three galaxies are all within 55$\arcsec$ of each other. In these cases, only two of the three galaxies in the triplet get measured redshifts. Naturally, higher order collision groups are also not fully recovered. 

Second, for those few galaxies in collision triplets and higher multiplicity collision groups that do not get measured redshifts, we use the nearest-neighbor approximation and assign the collided galaxies the redshifts of their nearest neighbors on the sky. By studying close pairs that have been recovered in the plate overlap regions, we find that this correction works well roughly two-thirds of the time for SDSS Main galaxies. In other words, roughly two-thirds of the time, two galaxies that are closer than 55$\arcsec$ on the sky are actually at the same redshift.

Third, we use the angular correlation function instead of the projected correlation function. This has the advantage that the angular separation of a given galaxy pair is completely unaffected by collisions, while the projected physical separation is not free of error. Since we are using angular instead of physical scales, a collided galaxy will only cause errors in our measurements if the nearest-neighbor redshift assignment causes the galaxy to drop out of or enter into the volume-limited sample in question. Errors in redshift that do not change a galaxy's membership in the sample (whether it was in or out) have zero effect on our measurements. Even when a collided galaxy's membership changes, there are really only three cases of collisions that can cause an error in our angular clustering measurements. The first occurs when the two galaxies straddle the redshift boundary of the volume-limited sample -- i.e., one is inside the sample and the other is outside the sample -- and the galaxy that is outside the sample did not get a measured redshift. The collided galaxy is then given the redshift of its neighbor and is thus brought into the sample. This error results in a small-scale pair that should not have been counted. The second failure mode occurs when the two collided galaxies straddle the luminosity limit of the sample, with the less distant galaxy being both outside the sample (i.e., below the luminosity limit) and the one that did not get a measured redshift. The collided galaxy is then moved to a larger distance and its calculated luminosity can now be high enough to bring it into the sample. This error also results in a small-scale pair that should not have been counted. The third and final failure mode occurs when the two collided galaxies are both in the sample, but the more distant one is close to the luminosity limit of the sample and does not get a measured redshift. When this galaxy is given the lower redshift of its neighbor, its calculated luminosity can make it drop out of the sample. This error results in a loss of a small-scale pair that should have been counted. All other cases of collisions result in no net gain or loss of a small-scale pair.

As a result of our methodology, only a very small fraction of all SDSS fiber collisions appear in our samples and only a small fraction of those that do appear actually cause errors on our measurements. Nevertheless, this small number of collided galaxies that do cause errors is not necessarily a negligible contribution to the number of galaxy pairs at very small scales. It is thus important to assess the magnitude of fiber collision errors in our analysis. One way to do that would be to use realistic mock galaxy catalogs that include fiber collisions so that we could directly test the extent to which our analysis method minimizes these errors. However, to be suitable for this purpose, the mock catalogs would have to cover the full flux-limited SDSS sample, making them a significant challenge to construct. Instead, we use a cross-correlation test that is designed to maximize the fiber collision signal present in our samples.

%-------------------------------
\begin{figure}
\centering
\includegraphics[width=3.2in,height=3.2in]{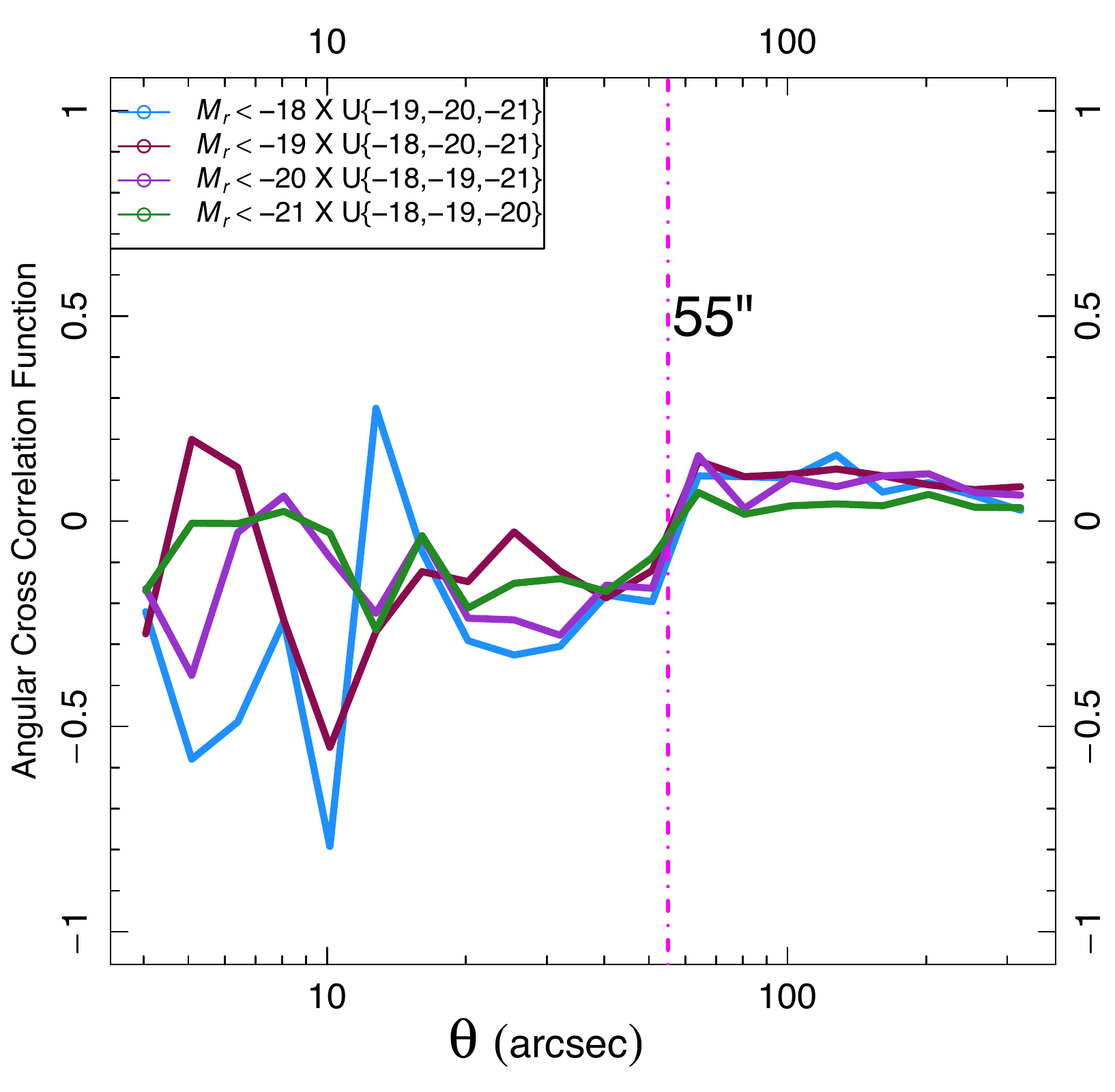} 
\caption{Cross-correlation test designed to probe the presence of fiber collision incompleteness in our volume-limited samples. The plot shows the angular cross-correlation function between sets of two samples that are designed to be spatially exclusive with each other. This is done by first introducing a low redshift cut in each of our four volume-limited samples so that no sample overlaps with another. We then measure the cross-correlation of each resulting sample with the union of the other three. With this setup there should be no real physical pairs in the cross-correlation except for erroneous pairs created due to fiber collisions. The 55$\arcsec$ collision scale is denoted by the vertical dashed line.
}
\label{fibcol_test}
\vspace{0.5cm}
\end{figure}
%-------------------------------

We first introduce a new low-redshift cut in each of our four volume-limited samples so that it does not overlap with any of the other samples. For example, we cut the $M_r<$ -21 sample at the outer redshift limit of the $M_r<$ -20 sample, we cut the $M_r<$ -20 sample at the outer redshift limit of the $M_r<$ -19 sample, and so forth. This results in four new volume-limited samples that have no spatial overlap with each other. We then measure the angular cross-correlation function between each of these new samples and the union of the other three new samples. In this way, each cross-correlation is measured between two samples that have no physical overlap and thus a minimal number of real physical correlated pairs, resulting in a measured cross-correlation that should be close to zero (it should actually be slightly higher than zero because there will be some real physical correlated pairs that straddle the redshift boundary between the samples). However, fiber collision errors can cause a fake signal in this cross-correlation because these errors can move galaxies across sample boundaries, as we have discussed above. Furthermore, this signal should only appear below the collision scale of 55$\arcsec$.

The results of this test are shown in Figure~\ref{fibcol_test}. The Figure shows exactly what we expect if fiber collision incompleteness is present in our samples. All four cross-correlations are slightly higher than zero for $\theta>55\arcsec$, and they exhibit a sharp decrement below this scale. The transition occurs exactly where we expect it, between the two bins that straddle the collision scale. Moreover, the suppression of the cross-correlation function below this scale appears to be roughly scale-independent, at least as far as we can tell with the precision level of the measurements. Finally, the fiber collision signal is similar for all four samples, with perhaps some slight suggestion of a larger effect for less luminous samples.

The cross-correlation test proves definitively that our analysis methodology does not eliminate fiber collision incompleteness. This incompleteness is clearly present in the angular clustering of our samples. However, it is very difficult to translate the signal in this cross-correlation test into an estimate of the effect of fiber collisions on our auto-correlation measurements shown in Figures~\ref{4-plot} and~\ref{HODfourplot}. The cross-correlation test maximizes the visible effect of fiber collisions in two ways. First, by removing any spatial overlap of the samples being cross-correlated, the unphysical pairs caused by fiber collisions become the entire signal, whereas in the auto-correlation measurements they are only a tiny fraction of the signal. Second, fiber collisions can only cause a deficit of pairs in this cross-correlation test, whereas they can both add or subtract pairs in the auto-correlation function, as we argued above. This adding and subtracting of pairs in the auto-correlation function could result in a smaller net effect due to fiber collisions. In the cross-correlation test, the only case of collisions that affects the measurement is when the two collided galaxies straddle the redshift boundary between the two samples. Regardless of which of these galaxies receives a redshift, the nearest-neighbor correction results in both galaxies ending up in the same sample, which removes the pair from the cross-correlation. This is why we see a deficit below the collision scale in Figure~\ref{fibcol_test}. The set of collisions affecting the cross-correlations is thus quite different from that affecting the auto-correlations, making it difficult to translate between the two.

We interpret the lack of any obvious visible features at the collision scale in our auto-correlation measurements as evidence that fiber collision errors in these measurements must be small relative to the real physical signal. Figure~\ref{fibcol_test} demonstrates that these features should appear between the two bins that straddle 55$\arcsec$, exactly as we expect. The various discontinuities seen in Figure~\ref{4-plot} occur at other scales and are thus not likely caused by fiber collisions. The only exception to this is the slight discontinuity seen in the auto-correlation of the $M_r<$-19 sample. However, the magnitude of this discontinuity is consistent with the up and down fluctuations seen at other scales. Moreover, there does not appear to be any systematic enhancement or suppression at all scales less than 55$\arcsec$, as is seen in the cross-correlation test. Errors due to fiber collisions are thus likely small relative to the real physical signals present in our auto-correlation measurements. Nevertheless, we emphasize that fiber collision errors are definitely present in our measurements at some level and it is not easy to asses their exact impact on our modeling results. Though we do not expect this impact to be large, it is prudent to treat our derived parameter values with some degree of caution.

%\bibliography{Biblio}

%================================================

\end{document}